\documentclass[draftclsnofoot,12pt,onecolumn]{IEEEtran}
\IEEEoverridecommandlockouts
\usepackage{cite}
\usepackage{amsmath,amssymb,amsfonts}
\usepackage{epstopdf,epsfig,psfrag,url,cite,multirow,float}
\usepackage{algorithm}
\usepackage{algorithmic}
\usepackage{graphicx}
\usepackage{textcomp}
\usepackage{xcolor}
\usepackage[small,bf]{caption}
\usepackage[font=footnotesize]{subfig}

\newcommand{\singlesize}{0.5} 
\newcommand{\doublesize}{0.48}

\newcommand{\bm}[1]{\boldsymbol{#1}}

\def\BibTeX{{\rm B\kern-.05em{\sc i\kern-.025em b}\kern-.08em
    T\kern-.1667em\lower.7ex\hbox{E}\kern-.125emX}}
\begin{document}

\title{Hierarchical Deep Reinforcement Learning for Age-of-Information Minimization in IRS-aided and Wireless-powered Wireless Networks}
\author{Shimin Gong, Leiyang Cui, Bo Gu, Bin Lyu, Dinh Thai Hoang, and Dusit Niyato\\
\thanks{Shimin Gong, Leiyang Cui, and Bo Gu are with the School of Intelligent Systems Engineering, Sun Yat-sen University, Shenzhen 518055, China (email: gongshm5@mail.sysu.edu.cn, cuily6@mail2.sysu.edu.cn, gubo@mail.sysu.edu.cn). Bin Lyu is with Key Laboratory of Ministry of Education in Broadband Wireless Communication and Sensor Network Technology, Nanjing University of Posts and Telecommunications, China (e-mail: blyu@njupt.edu.cn). Dinh Thai Hoang is with the School of Electrical and Data Engineering, University of Technology Sydney, Australia (email: hoang.dinh@uts.edu.au). Dusit Niyato is with School of Computer Science and Engineering, Nanyang Technological University, Singapore (email: dniyato@ntu.edu.sg).}


}
\maketitle

\begin{abstract}
In this paper, we focus on a wireless-powered sensor network coordinated by a multi-antenna access point (AP). Each node can generate sensing information and report the latest information to the AP using the energy harvested from the AP's signal beamforming. {{blue}We aim to minimize the average age-of-information (AoI) by adapting the nodes' transmission scheduling and the transmission control strategies jointly}. To reduce the transmission delay, an intelligent reflecting surface (IRS) is used to enhance the channel conditions by controlling the AP's beamforming {{blue}vector} and the IRS's phase shifting {{blue}matrix}. {{blue}Considering dynamic data arrivals at different sensing nodes}, we propose a hierarchical deep reinforcement learning (DRL) framework to {{blue}for AoI minimization in two steps}. The users' transmission scheduling is firstly determined by the outer-loop DRL approach, e.g.~the DQN or PPO algorithm, and then the inner-loop optimization is used to adapt either the uplink information transmission or downlink energy transfer to all nodes. A simple and efficient approximation is also proposed to reduce the {{blue}inner-loop} rum time overhead. Numerical results verify that the hierarchical learning framework outperforms typical baselines in terms of the average AoI and proportional fairness among different nodes.
\end{abstract}


\section{Introduction}
With the development of the future Internet of Things (IoT), a large portion of the emerging applications (e.g., autonomous driving, interactive gaming, and virtual reality) depends on timely {transmissions and processing of the IoT devices' sensing data in the physical world, e.g.,~\cite{aoi-role} and~\cite{aoi-dusit}.} Maintaining information freshness in time-sensitive applications requires a new network performance metric, i.e.,~the age-of-information (AoI), which is defined as the elapsed time since {the generation of the latest status update successfully received by the receiver}~\cite{aoi-tit}. In wireless networks, the overall time delay of each node is mainly caused by the waiting time or scheduling delay before information transmission and the in-the-air transmission delay. The waiting delay is usually determined by the multi-user scheduling strategy, while the transmission delay depends on the {network capacity and channel conditions, e.g.,~mutual interference or channel fading effect}. To minimize the transmission delay, we need to explore preferable channel opportunities and adapt the transmission control parameters accordingly, e.g.,~power control, channel allocation, and beamforming strategies. More recently, the intelligent reflecting surface (IRS) has been used to reduce the transmission delay by shaping the wireless channels in favor of information transmission via passive beamforming {optimization}~\cite{irs-tut}. The AoI-aware scheduling and transmission control in wireless networks {were previously} studied by the queueing theory~\cite{fcfs-lcfs}. The closed-form analysis of the AoI performance is tedious and {usually difficult}, typically relying on {specific probability distributions of the sensor nodes' data arrivals}, the data transmission/service time, and {the rate of information requests at} user devices. For more complex wireless networks, e.g., with {users' mobility and limited energy resources}, it is still {very challenging} to optimize multi-user scheduling policy {to maximize} the overall AoI performance.

In energy-constrained wireless networks, the AoI minimization depends on the optimal control of each user's energy supply and demand, especially in wireless powered communication networks. {When a sensor node is scheduled more often} to update its sensing information, more wireless energy transfer is required {for the} node {to achieve} self-sustainability. This {may reduce the transmission opportunities and the energy transfer to other nodes, leading to} the AoI-energy tradeoff study in energy-constrained wireless networks~\cite{cache-aoi}. The joint optimization of multi-user scheduling and energy management is {usually formulated as a dynamic program} due to the spatial-temporal couplings among wireless {sensor nodes}. With limited energy supply, the {users'} scheduling becomes more challenging to balance the tradeoff between AoI and energy consumption. The AoI minimization problem is further complicated by the {unknown channel} conditions and the sensor nodes' {dynamic data arrivals}. Without complete system information, the scheduling strategy has to be adapted according to the users' historical AoI information. Instead of the optimization approaches, the AoI minimization problems can be flexibly solved by the model-free deep reinforcement learning (DRL) approaches, e.g.,~\cite{cache-aoi,fas-dqn,harq-jsac,delay-target}. The DRL approaches can reformulate the AoI minimization problem into Markov decision process (MDP). The action includes the scheduling and transmission control strategies. With {a large-size IRS, the action and state spaces become high-dimensional and lead to unreliable and slow learning performance. This motivates us to design a more efficient learning approach for the AoI minimization problem.}

Specifically, in this paper, we aim to minimize the average AoI in a wireless-powered network, consisting of a multi-antenna access point (AP), a wall-mounted IRS, and multiple single-antenna sensor nodes, sampling the status information from different physical processes. The IRS is used to enhance the channel conditions and reduce the transmission delay. It can assist either the downlink wireless power transfer from the AP to {the sensing nodes} or uplink information transmissions from {nodes} to the AP~\cite{irs-tut}. The joint scheduling and beamforming optimization normally leads to a high-dimensional mix-integer problem that is difficult to solve practically. {Different from the conventional DRL solutions, we devise a two-step hierarchical learning framework to improve the overall learning efficiency. The basic idea is to adapt the scheduling strategy by the outer-loop DRL algorithm, e.g.,~the value-based deep Q-network (DQN) or the policy-based proximal policy optimization (PPO) algorithm~\cite{PPO}, and optimize the beamforming strategy by the inner-loop optimization module.} Given the outer-loop decision, the inner-loop procedures either optimize the AP's downlink energy transfer or optimize individual's uplink information transmission. Specifically, the contributions of this paper are summarized as follows:
\begin{itemize}
    \item {\emph{The IRS-assisted scheduling {and beamforming} for AoI minimization:}} We aim to reduce both the packet waiting and transmission delays for updating sensing information in a wireless-powered and IRS-assisted wireless network. {The IRS's passive beamforming not only enhances the wireless power transfer to sensor nods, but also assists their uplink information transmissions to the AP. We formulate the AoI minimization problem by jointly adapting the user scheduling and beamforming strategies.} 
    \item {\emph{The hierarchical DRL approach for {AoI minimization}:}} {A hierarchical DRL framework is proposed to solve the AoI minimization in two steps}. The model-free DRL in the outer loop adapts the combinatorial scheduling decision according to each user's energy status and the AoI performance. Given the outer-loop scheduling, we optimize the joint beamforming strategies for either the downlink energy transfer and the uplink information transmission.
    \item {\emph{{Policy-based PPO algorithm} for outer-loop scheduling:}} Our simulation results verify that the hierarchical learning framework significantly reduces the average AoI compared with typical baseline strategies. Besides, we compare both the traditional DQN and the PPO methods for the outer-loop scheduling optimization. The PPO-based hierarchical learning can improve {convergence and {achieve} a lower AoI value compared to the DQN-based method.} %
\end{itemize}

Some preliminary results of this work {have been presented in}~\cite{conf-version}. In this extension, we include detailed analysis about the PPO algorithm and compare it with the DQN method. We also propose a simple approximation for the inner-loop optimization to reduce the time overhead while achieving comparable AoI performance at convergence. The remainder of this paper is organized as follows. We discuss related works in Section~\ref{sec-related} and present our system model in Section~\ref{sec:model}. We present the hierarchical learning framework in Section~\ref{sec:learning}. The inner-loop optimization and outer-loop learning procedures are detailed in Sections~\ref{sec-innerloop} and~\ref{sec-outerloop}, respectively. Finally, we present the numerical results in Section~\ref{sec:num} and conclude the paper in Section~\ref{sec:cons}.

\section{Related works}\label{sec-related}


\subsection{DRL Approaches for AoI Minimization}


DRL has been introduced recently as an effective solution {for AoI minimization} by adapting the scheduling and beamforming strategies according to time-varying traffic demands and channel conditions. The authors in~\cite{fas-dqn} designed the freshness-aware scheduling solution by using the DQN method. The DQN agent continuously updates its scheduling {strategy to maximize the freshness of information} in the long term. {The authors in~\cite{harq-jsac} focused on a multi-user status update system, where a single sensor node monitors a physical process and schedules its information updates to multiple users with time-varying channel conditions.
Based on the user's instantaneous ACK/NACK feedback, the DRL agent at the information source can decide on when and to which user to transmit the next information update, without any priori information on the random channel conditions.} The authors in~\cite{delay-target} focused on a multi-access system where the base-station (BS) aims to maximize its information collected from all wireless users. Given a strict time deadline to each wireless user, {the PPO algorithm was used} to adapt the scheduling policy by learning the users' traffic patterns from the past experiences. The authors in~\cite{multi-sched} considered a different multi-user scheduling scheme that allows a group of sensor nodes to transmit their information simultaneously. The scheduling decision is adapted to minimize the AoI by {using the double DQN (DDQN) method~\cite{ddqn}, an extension of the DQN method by using two sets of deep neural networks (DNNs) to approximate the Q-value.}
The AoI-energy tradeoff study in~\cite{DRL_age_engrgy_tradeoff} revealed that the sensor nodes' energy consumption can be reduced without a significant increase in the AoI, by using DQN to adapt the content update in the caching node. 



\subsection{RF-powered Scheduling for AoI Minimization}
The wireless power transfer and energy harvesting are promising techniques to sustain the massive number of low-cost sensor nodes. {However, energy harvesting is usually unreliable depending on the channel conditions. The dynamics and scarcity in energy harvesting make it more challenging for the sensor nodes' energy management and AoI minimization. The authors in~\cite{aoi-crn} focused on AoI minimization in a cognitive radio network with dynamic supplies of the energy and spectrum resources. The optimal scheduling policy is derived by a dynamic programming approach, revealing a threshold structure depending on the sensor nodes' AoI states. The authors in~\cite{poor20} revealed that the optimal policy allows each sensor to send a status update only if the AoI value is higher than some threshold that depends on its energy level. Considering stochastic energy harvesting at sensor devices, the authors in~\cite{xuchen21} studied the AoI minimization problem with the long-term energy constraints and proposed Lyapunov-based dynamic optimization to derive an approximate solution.
Generally the dynamic optimization approaches are not only computational demanding, but also relying on the availability of system information. Without knowing the dynamic energy arrivals, the authors in~\cite{aoi-energy-q-learning} reformulated the AoI minimization problem into MDP. The online Q-learning method was proposed to adaptively schedule the wireless devices' information update.}
The authors in~\cite{rf-aoi} focused on the RF-powered wireless network, where the wireless devices can harvest RF power from a dedicated BS and then transmit their update packets to the BS. To minimize the long-term AoI, the DQN algorithm was used to adapt the scheduling between the downlink energy transfer and the uplink information transmission. {Both the DQN and dueling DDQN methods were used in~\cite{ddqn-aoi22} to adapt the sensing and information update policy for AoI minimization in a spectrum sharing cognitive radio system. Considering energy harvesting ad hoc networks, the authors in~\cite{A2C-aoi} solved the AoI minimization problem by using the advantage actor-critic (A2C) algorithm to adapt the scheduling and power allocation policy, which shows faster runtime and comparable AoI performance to the optimum.}
{The above-mentioned works typically solve the AoI minimization by using the conventional model-free DRL methods. These methods become inflexible and unreliable due to slow convergence with the increasing state and action spaces.}

\subsection{IRS-Assisted AoI Minimization}
The IRS's reconfigurability can be used to enhance the channel {quality/capacity} or reduce the transmission delay by tuning the phase shifts of the reflecting elements~\cite{irs-tut}. {Only a few existing works have discussed the IRS's application for AoI minimization in wireless networks.} The authors in~\cite{min-delay-jsac} {focused on a mobile edge computing (MEC) system and proposed using the IRS to minimize} the workload processing delay by optimizing the IRS's passive beamforming strategy. The authors in~\cite{delay-con} set delay constraints to the wireless users' uplink information transmissions and revealed that the IRS's passive beamforming can help reduce the wireless users' transmit power. The authors in~\cite{irs-aoi-arxiv} employed the IRS to enhance the {AoI performance by jointly optimizing the uses' scheduling} and the IRS's passive beamforming strategies. The combinatorial scheduling decision is adapted by the model-free DRL algorithm, while the passive beamforming optimization relies on the solution to the conventional semi-definite relaxation (SDR) problem. The authors in~\cite{irs-aoi-tvt} employed the UAV-carrying IRS to relay information from the ground users to the BS. The AoI minimization requires the optimization of the UAV's altitude, the ground users' transmission scheduling, and the IRS's passive beamforming strategies. Comparing to~\cite{irs-aoi-arxiv} and~\cite{irs-aoi-tvt}, our work in this paper exploits the performance gains in both the uplink and downlink of the IRS-assisted system. {The IRS's passive beamforming} not only assists uplink information transmission but also enhances or balances the AP's downlink energy transfer to the users.


\section{System model}\label{sec:model}

We consider an IRS-assisted wireless sensor network deployed in smart cities to assist information sensing and decision making, similar to that in~\cite{conf-version}. The system consists of a multi-antenna AP with $M$ antennas, an IRS with $N$ reflection elements, and $K$ single-antenna IoT devices, denoted by the set $\mathcal{K}\triangleq\{1,2,\ldots,K\}$. The system model can be {straightforwardly} extended to the cases with multiple AP or multiple IRSs.
The sensing information is typically a small amount of data, which should be timely updated to the AP for real-time status monitoring. We assume that all sensor nodes are low-power devices and wireless powered by harvesting RF energy from the AP's beamforming signals. {The wireless powered communications technology has been verified and evaluated in~\cite{lora22}, showing that the LoRa-based sensor nodes typically have 0.5-1.5 mJ energy consumption, and require 2-5 seconds of energy harvesting time 3.2 meters away to sustain periodical information sensing and data transmissions up to 200 bytes.}
The IRS can be deployed on the exterior walls of buildings to enhance the channel conditions between the sensor nodes and the AP. We aim to collect all sensor nodes' data {in a timely fashion} by scheduling their uplink data transmissions, based on their channel conditions, traffic demands, and energy status. {A list of notations is provided in Table~\ref{tab-notation}.}

\begin{table}[]
    \centering
    \caption{A list of Notations}
    \setlength{\tabcolsep}{2.8mm}{
    \begin{tabular}{|c|c||c|c|}
    \hline
    \makebox[0.1\textwidth][c]{Notation} & \makebox[0.2\textwidth][c]{Description} & \makebox[0.1\textwidth][c]{Notation} & \makebox[0.2\textwidth][c]{Description} \\ \hline
    $M$ & Number of AP's antennas & $N$ & Size of the IRS\\  \hline
    $K$ & Number of IoT devices & $\mathcal{T}$ & The set of time slots \\  \hline
    $\psi_0(t)\in\{0,1\}$ & The AP's mode selection & $\psi_k(t)\in\{0,1\}$ & The AP's uplink scheduling\\  \hline
    ${\bf G}(t)$ & The AP-IRS channel matrix  & ${\bf h}^{{r}}_{k}(t)$ & The IRS-User channel vector \\  \hline
    ${\bm{\Theta}}_d(t)$, ${\bm{\Theta}}_u(t)$ & The IRS's beamforming strategies & ${\bf w}_d(t)$, ${\bf w}_u(t)$ & The AP's beamforming vectors\\  \hline
    $\eta$ &  Energy conversion efficiency & $E^{h}_{k}(t)$ & Energy harvested by the user-$k$ \\ \hline
    $E_{k}^c(t)$ & The user-$k$'s energy consumption & ${E}_k(t)$ & The user-$k$'s energy state \\  \hline
    $E_{\max}$ & Maximum energy capacity & $\gamma_k(t)$ & The received SNR at the AP\\  \hline
    $r_k(t)$ & The user-$k$'s uplink throughput & $\tau_k$ & The user-$k$'s uplink transmission time\\  \hline
    $d_k$ & The user-$k$'s data size &  $A_k(t)$ & The user-$k$'s AoI value \\ \hline \end{tabular}}\label{tab-notation}
    \vspace{-0.7cm}
\end{table}

\subsection{Mode Selection and Scheduling}

We consider a time-slotted frame structure to avoid contention between different nodes. Each data frame is equally divided into $T$ time slots allocated to different sensor nodes. Let $\mathcal{T}\triangleq\{1,2,\ldots, T\}$ denote the set of all time slots. In each time slot, we need to firstly decide the AP's operation mode, i.e., the time slot is used for either the downlink energy transfer or the uplink data transmission.
We use $\psi_0(t)\in\{0,1\}$ to denote the AP's mode selection in each time slot, i.e., $\psi_0(t)=1$ indicates the downlink energy beamforming and $\psi_0(t)=0$ represents the uplink information transmission. We further use $\psi_k(t)\in\{0,1\}$ to denote the uplink scheduling policy, i.e., $\psi_k(t)=1$ represents that the $k$-th sensor node is allowed to access the channel for uploading its sensing information to the AP. We require that at most one sensor node can access the channel in each time slot, which implies the following scheduling constraint:
\begin{equation}\label{equ-sched}
\psi_0(t) + \sum_{k\in\mathcal{K}}\psi_k(t) \leq 1, \quad \forall\, t \in \mathcal{T}.
\end{equation}
We denote $\bm{\Psi}(t)=[\psi_0(t), \psi_1(t), \ldots, \psi_K(t)]$ as the AP's scheduling policy, which depends on the sensor nodes' traffic demands, channel conditions, and energy status. 

%

Let $\mathcal{N}\triangleq \{1,2,\ldots,N\}$ denote the set of the IRS's reflecting elements and $\theta_{n}(t) \in (0, 2\pi]$ denote the phase shift of the $n$-th reflecting element in the $t$-th time slot. We define the IRS's phase shifting vector in the $t$-th time slot as $\bm{\theta}(t) =[e^{j\theta_{n}(t)}]_{n\in\mathcal{N}}$. Note that the IRS can set different beamforming vectors, denoted as $\bm{\theta}_{d}(t)\triangleq[e^{j\theta_{d,n}(t)}]_{n\in\mathcal{N}}$ and $\bm{\theta}_u(t)\triangleq[e^{j\theta_{u,1}(t)}]_{n\in\mathcal{N}}$, for the downlink and uplink phases, respectively. The channel matrix from the AP to the IRS in $t$-th time slot is given by ${\bf G}(t) \in \mathbb{C}^{M \times N}$. The channel vectors from the IRS and the AP to the $k$-th sensor node are denoted by ${\bf h}^{{r}}_{k}(t)\in \mathbb{C}^{N \times 1}$ and ${\bf h}^{{d}}_{k}(t)\in \mathbb{C}^{M \times 1}$, respectively.
The AP can estimate the channel information by a training period {at} the beginning of each time slot.

\subsection{Downlink Energy Transfer}
When $\psi_0(t) = 1$, the IRS-assisted downlink energy transfer ensures the sustainable operation of the system. Given the IRS's passive beamforming strategy ${\bm{\theta}}_d(t)$, the equivalent downlink channel vector ${\bf f}_{d,k}(t)$ from the AP to the $k$-th sensor node can be expressed as follows:
\begin{equation}\label{channel_down}
{\bf f}_{d,k}(t) = {\bf h}^{{d}}_{k}(t) + {\bf G}_{k}(t){\bm{\Theta}}_d(t){\bf h}^{{r}}_{k}(t),
\end{equation}
where we define $\bm{\Theta}_d(t) \triangleq\text{diag}(\bm{\theta}_d(t))$ as a diagonal matrix with the diagonal element $\bm{\theta}_d(t)$. The phase shifting matrix $\bm{\Theta}_d(t)$ represents the IRS's passive beamforming strategy in the downlink energy transfer. 
Let ${\bf w}_d(t) \in \mathbb{C}^{M \times 1}$ denote the AP's transmit beamforming vector in the downlink energy transfer phase. Given the AP's transmit power $p_s$, the AP's beamforming signal is given by ${\bf x}(t) = \sqrt{p_s}{\bf w}_d(t) s_0(t) $, where $s_0(t)\in\mathbb{C}$ denotes a random complex symbol with the unit power. Then, the received signal at the $k$-th sensor node is given as ${y}_k(t) = {\bf f}^{{H}}_{d,k}(t) {\bf x}(t) + n_k(t)$, where $(\cdot)^H$ denotes conjugate transpose and $n_k(t)$ is the {normalized Gaussian noise with zero mean and unit power}. Considering a linear energy harvesting (EH) model~\cite{en_linear_model_2}, the received signal ${y}_k(t)$ can be converted to energy as follows:
\begin{equation}\label{equ-eh}
E^{h}_{k}(t) =\eta \mathbb{E}[ | {\bf f}^{{H}}_{d,k}(t) {\bf x}(t)|^2]  =  \eta p_s | {\bf f}^{{H}}_{d,k}(t) {\bf w}_d(t)|^2 ,
\end{equation}
where $\eta $ denotes the energy conversion efficiency. The energy harvested from the noise signal is assumed to be negligible. It is clear that the AP can control the energy transfer to different sensor nodes by optimizing the downlink beamforming vector ${\bf w}_d(t)$.

In particular, the AP can steer the beam direction toward the sensor nodes with insufficient energy supply. Besides, the IRS's passive beamforming strategy $\bm{\Theta}_d(t)$ controls the downlink channel conditions ${\bf f}_{d,k}(t)$ to individual receivers. Due to the broadcast nature of wireless signals, the AP's energy transfer to different sensor nodes {depends on} the joint beamforming strategies $({\bf w}_d(t) ,\bm{\Theta}_d(t))$ in different time slots. A more practical non-linear EH model can be also applied to our system. In this case, the harvested power firstly increases with the received signal power and then becomes saturated as the received signal power continues to increase, e.g.,~\cite{non_linear_eh_2}. This can be approximated by a piecewise linear EH model: $E^{h}_{k}(t) = \min \Big\{ \eta p_s | {\bf f}^{{H}}_{d,k}(t) {\bf w}_d(t)|^2, \,\, p_{\text{sat}}\Big\}$, where $p_{\text{sat}}$ denotes the saturation power. Our algorithm in the following part adopts the linear EH model in~\eqref{equ-eh} and can be easily applied to the piecewise linear model with minor modifications.


\subsection{Sensing Information Updates}
We assume that the uplink channels are the same as the downlink channels {in each time slot due to channel reciprocity, similar to that in~\cite{conf-version} and~\cite{rf-aoi}}. Let $p_k(t)$ denote the transmit power of the $k$-th sensor node when it is scheduled in the $t$-th time slot, i.e., $\psi_k(t) = 1$. The signal received at the AP is given by $\textbf{y}_{k} = \sqrt{p_k(t)}{\bf f}_{u,k}(t) s_{k} + {\bf n}_{k}(t)$, where ${\bf f}_{u,k}(t)$ denotes the uplink channel from {the $k$-th sensor node to the AP and $s_{k}(t)$ denotes its} information symbol. Similar to~\eqref{channel_down}, {the IRS-assisted} uplink channel is given by ${\bf f}_{u,k}(t) = {\bf h}^{{d}}_{k}(t) + {\bf G}_{k}(t){\bm{\Theta}}_u(t){\bf h}^{{r}}_{k}(t)$, where ${\bm{\Theta}}_u(t)\triangleq\text{diag}(\bm{\theta}_u(t))$ denotes the IRS's uplink passive beamforming strategy. Without loss of generality, the noise signal ${\bf n}_{k}(t)$ received by the AP can be normalized to the unit power. {Thus, the} received SNR can be characterized as $\gamma_k(t) = p_k(t) |{\bf f}_{u,k}^H(t){\bf w}_u(t)|^2$, where ${\bf w}_u(t)$ represents the AP's receive {beamforming vector}. By using the time division protocol, the sensor nodes' uplink transmissions can avoid mutual interference. {The AP can simply align its receive beamforming vector} ${\bf w}_u(t)$ to the uplink channel ${\bf f}_{u,k}(t)$. As such, we can denote the received SNR as $\gamma_k(t) = p_k(t) ||{\bf f}_{u,k}(t)||^2$ and characterize the uplink throughput {as $r_k(t) = \tau_k\log \bigl(1+p_k(t) ||{\bf f}_{u,k}(t)||^2 \bigr)$, where $\tau_k\in[0,1]$ denotes the uplink transmission time. Given the data size $d_k$}, we require $r_k(t) \geq d_k$ to ensure the successful transmission of the sensing information.


In each time slot, the sensor node's energy consumption is given by~$E_{k}^c(t) = \tau_k(t) (p_k(t) + p_c)$, where $p_c$ denotes a constant circuit power to maintain the node's activity. The power consumption $\tau_k(t) p_k(t)$ in uplink data transmission is linearly proportional to the transmit power $p_k(t)$ and the transmission time $\tau_k$. The transmit power $p_k(t)$ can vary with the channel conditions {to ensure} the rate constraint $r_k(t) \geq d_k$. Let $E_{\max}$ denote the sensor nodes' maximum battery capacity and ${E}_k(t)$ be the energy state in the $t$-th time slot. Then, we have the following energy dynamics:
\begin{equation}\label{equ-energy-dyna}
{E}_k(t+1)= \min\Big\{ \Big ({E}_k(t) -\psi_{k}(t)E^c_{k}(t) \Big)^+ + \psi_{0}(t)E^h_{k}, E_{\max}\Big\}.
\end{equation}
Here we denote $(x)^+\triangleq\max\{x, 0\}$ for simplicity. 

\section{Hierarchical Learning for AoI Minimization}\label{sec:learning}

{In this paper, the physical sensing process is beyond our control and we only focus on the transmission scheduling and beamforming optimization over the wireless network. The sensing information can be randomly generated by the sensor nodes, depending on the energy status and the physical process under monitoring.}
Once new sensing data arrives, each sensor node will discard existing data in the cache and always cache the latest sensing data. {From the perspective of the receiver, the sensing data from each sensor node is considered as the new information and used to replace the obsolete information at the receiver.} When the node-$k$ is scheduled for uplink data transmission, the cached information will be uploaded to the AP and then the AP will replace the sensing information by the latest copy.

For each sensor node $k\in\mathcal{K}$, the caching delay depends on the AP's scheduling policy $\bm{\Psi}(t)$.
The transmission delay can be minimized by optimizing the sensor node's transmit control strategy $(p_k(t),\tau_k(t))$ and the joint beamforming strategies $({\bf w}_u(t), {\bm \Theta}_u(t))$ in the uplink transmissions. Let $A_k(t)$ denote the AoI value of the $k$-th sensor node, which is used to characterize the information freshness at the AP. When the node-$k$ is scheduled to update its information in the $t$-th time slot, the AP can replace the obsolete information by the new sensing information and thus update the AoI in the next time slot as $A_k(t+1) = 1$. Here we assume that the node-$k$ can successfully finish its data transmission at the end of each time slot. Otherwise, when the node-$k$ is not scheduled, its AoI will be further increased by one, i.e., $A_k(t+1) = A_k(t)+ 1$. Therefore, the AoI of each sensor node $k\in\mathcal{K}$ will be updated by the following rules:
\begin{equation}\label{equ-aoi-dyna}
A_k(t+1) =
\begin{cases}
1, & \text{if } o_k(t)=1, \psi_k(t)=1,  r_k(t) \geq d_k, E_k(t)\geq E^c_{k}(t), \\
A_k(t) + 1, & \text{otherwise}.
\end{cases}
\end{equation}
Here $o_k(t)\in\{0,1\}$ indicates the status of the caching space. When the cache is non-empty with $o_k(t) =1$ and the node-$k$ is currently scheduled with $\psi_k(t) = 1$, the AP can update the sensing information if the uplink data transmission is successful. Given the size $d_k$ of sensing data, the scheduled node-$k$ {will have a successful data transmission if it has sufficient energy}, i.e., $E_k(t)\geq E^c_{k}(t)$, to fulfill its traffic demand, i.e.,~$r_k(t) \geq d_k$, where the uplink data rate $r_k(t)$ depends on the control parameters $(p_k(t),\tau_k)$ and the joint beamforming strategies $({\bf w}_u(t), {\bm \Theta}_u(t))$.

We aim to minimize the AoI by optimizing the scheduling policy $\bm{\Psi}\triangleq\{\bm{\Psi}(t)\}_{t\in\mathcal{T}}$ and the joint beamforming strategies $({\bf w}, {\bm \Theta}) \triangleq ({\bf w}_m(t), {\bm \Theta}_m(t))_{m\in\{d,u\}, t\in\mathcal{T}}$ in both the downlink and uplink phases.
Considering different priorities of the sensing information, we assign different weights to the AoI values and define the time-averaged weighted AoI as follows:
\begin{equation}\label{equ-waoi}
\bar{A}(\bm{\Psi},{\bf w},\bm{\Theta}) = \lim_{T\rightarrow \infty}\frac{1}{TK}\mathbb{E}\left[\sum_{t\in\mathcal{T}}\sum_{k\in\mathcal{K}}\lambda_kA_k(t)\right],
\end{equation}
where the constant $\lambda_k$ indicates the delay sensitivity of different sensing information. A larger weight should be given to more critical sensing information, e.g.,~the safety monitoring in autonomous driving. Till this point, we can formulate the AoI minimization problem as follows:
\begin{align}\label{prob-aoi}
\min_{\bm{\Psi},{\bf w},\bm{\Theta}} ~&~\bar{A}(\bm{\Psi},{\bf w},\bm{\Theta}),
\quad \text{s.t.} \,\, \eqref{equ-sched}-\eqref{equ-aoi-dyna}. 
\end{align}
Given the mode selection $\psi_0(t)$, the optimization of $({\bf w},\bm{\Theta})$ corresponds to either the downlink energy transfer or the uplink information transmission. The downlink energy transfer determines the power budgets of different sensor nodes, which should be jointly optimized with the users' scheduling policy $\{\psi_k(t)\}_{k\in\mathcal{K}}$ to improve the overall AoI performance. The problem~\eqref{prob-aoi} is firstly challenged by the stochasticity {and high-dimensionality}. The energy dynamics in~\eqref{equ-energy-dyna} and the time-averaged AoI objective in~\eqref{prob-aoi} imply that the solutions $(\bm{\Psi}(t),{\bf w}(t), \bm{\Theta}(t))$ in each time slot are temporally correlated. A dynamic programming approach to solve~\eqref{prob-aoi} can be practically intractable due to the curse of dimensionality. {The joint scheduling and beamforming optimization also lead to a high-dimensional mix-integer problem that is difficult to solve practically}. The second difficulty of problem~\eqref{prob-aoi} lies in that the dynamics of the data arrival process at each sensor node can be unknown to the AP, which makes the scheduling optimization more complicated in practice.
Without complete information, the AP has to adapt its scheduling policy based on the past observations of the AoI dynamics.
The third difficulty lies in the combinatorial nature of the AP's scheduling policy $\bm{\Psi}$. Given the scheduling policy $\bm{\Psi}$, the challenges still exist as the joint beamforming strategies $({\bf w}, \bm{\Theta})$ are not only coupled with each other in a non-convex form, but also introduce the competition for energy resource among different sensor nodes. 

\begin{figure}[t]
	\centering
	\includegraphics[width=\singlesize\textwidth]{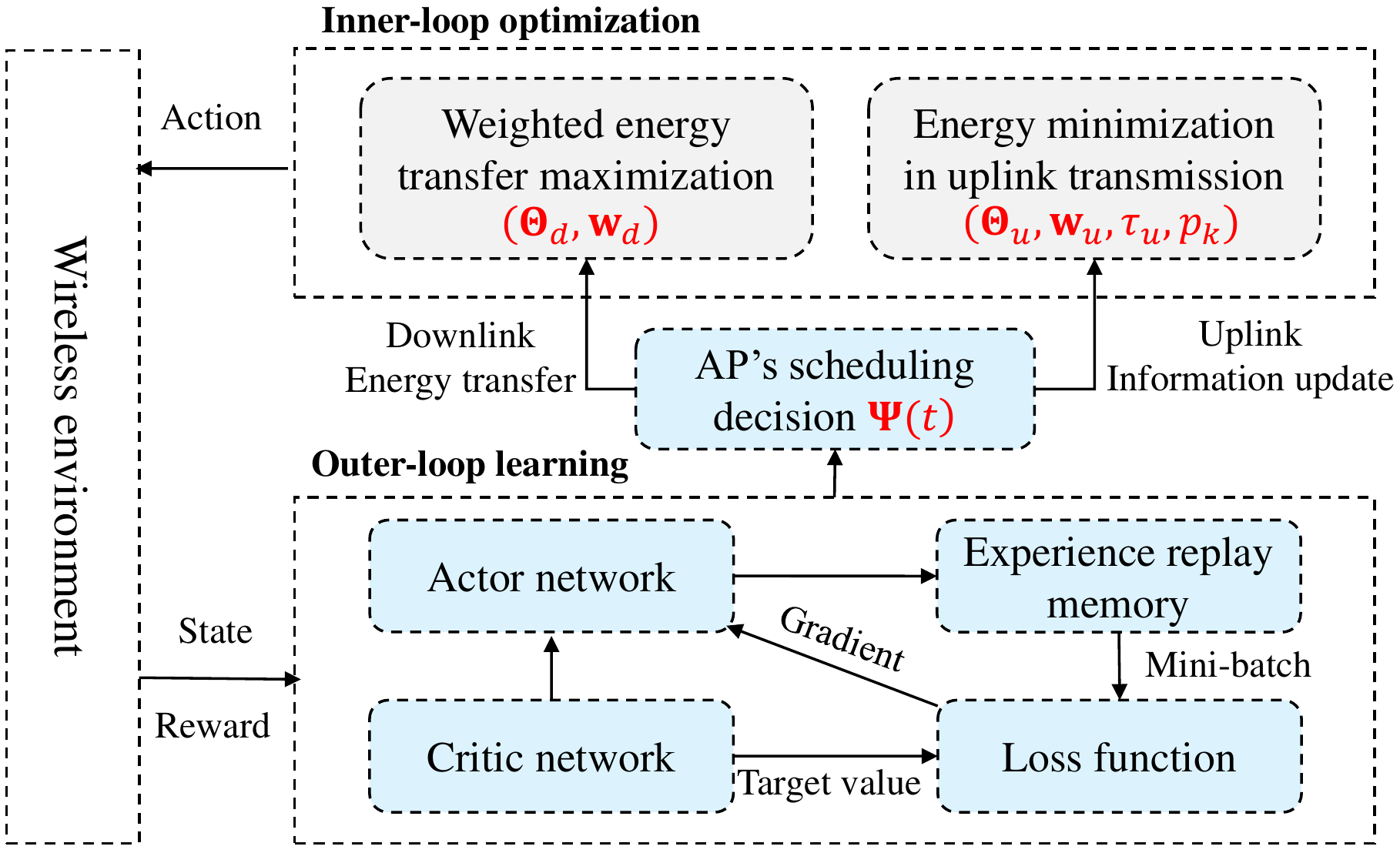}
	\caption{The hierarchical DRL framework includes the outer-loop DRL and the inner-loop optimization methods.}\label{fig-algo}
\vspace{-1cm}
\end{figure}

To {overcome} these difficulties, we devise a hierarchical learning structure for problem~\eqref{prob-aoi} that decomposes the optimization of $(\bm{\Psi},{\bf w}, \bm{\Theta})$ into two parts. The overall algorithm framework is shown in Fig.~\ref{fig-algo}, which mainly includes the outer-loop learning module for scheduling and the inner-loop optimization module for beamforming optimization. {In fact, we may also apply an inner-loop learning method to adapt the beamforming strategy. However, this may still require huge action and state spaces and thus lead to slow learning performance. Instead of an inner-loop learning method, the AP can estimate the beamforming strategy $({\bf w}(t), \bm{\Theta}(t))$ efficiently by using the optimization method, based on the AP's observation of the users' channel conditions. This motivates us to devise a hybrid solution structure that exploits the outer-loop learning and the inner-loop optimization modules. Specifically,} due to the combinatorial nature of the scheduling policy $\bm{\Psi}(t)$, we employ the model-free DRL approach to {adapt} $\bm{\Psi}(t)$ in the outer-loop learning procedure. In each iteration, the DRL agent firstly determines $\bm{\Psi}(t)$ based on the past observations of the nodes' AoI dynamics. Then, the inner-loop joint optimization of $({\bf w}(t), \bm{\Theta}(t))$ becomes much easier by using the alternating optimization (AO) and semi-definite relaxation (SDR) methods. According to the outer-loop mode selection $\psi_0(t)$, the inner-loop optimization aims to either maximize the downlink energy transfer to all sensor nodes or fulfill the uplink information transmission of the scheduled sensor node. After the inner-loop optimization, the AP can execute the joint action $(\bm{\Psi}(t),{\bf w}(t), \bm{\Theta}(t))$ in the $t$-th time slot and then update the AoI state of each sensor node. The evaluation of the time-averaged AoI performance will drive the outer-loop DRL agent to adapt the scheduling decision $\bm{\Psi}(t+1)$ and the beamforming strategies $({\bf w}(t+1), \bm{\Theta}(t+1))$ in the next time slot. By such a decomposition, the inner-loop optimization becomes computation-efficient, while the outer-loop learning becomes time-efficient as it only adapts the combinatorial scheduling policy with a smaller action space.


\section{Inner-loop Optimization Problems}\label{sec-innerloop}

Given the scheduling decision $\psi_0(t)\in\{0,1\}$, the AP either beamforms RF signals for downlink energy transfer or receives the sensing information from individual sensor node. In each case, the AP will optimize the joint beamforming strategies $({\bf w}(t),\bm{\Theta}(t))$. In the sequel, we discuss the inner-loop optimization problems in two cases.

\subsection{Energy Minimization in Uplink Transmission}\label{subsec-uplink}
In the $t$-th time slot, when the $k$-th sensor node is allowed to update its sensing information with $\psi_k(t) = 1$, all other sensor nodes have to wait {for scheduling in the other time slots}. The sensor nodes' AoI values will be either increased by 1 or reset to 1 if the transmission is unsuccessful or successful, respectively, as shown in~\eqref{equ-aoi-dyna}. In this case, we can minimize the energy consumption $E^c_{k}(t) = \tau_k(t)(p_k(t) + p_c)$ of the scheduled sensor node conditioned on the successful transmission of its sensing data, i.e.,~$r_k(t) \geq d_k$. This will preserve more energy for its future use. Thus, we have the following energy minimization problem:
\begin{subequations}\label{prob-uplink-energy}
\begin{align}
\min_{\tau_k,p_k,{\bf w}_u,\bm{\Theta}_u} ~&~ \tau_k(p_k  + p_c)\label{obj-energy-ul}\\
\text{s.t.}
~&~ \tau_k \log(1+ p_k |{\bf f}^H_{k} {\bf w}_u|^2) \geq d_k, \label{con-rate-ul}\\
~&~ \tau_k \in(0,1)\text{ and }\theta_{u,n} \in (0, 2\pi], \, n\in\mathcal{N}.
\end{align}
\end{subequations}
The uplink transmission only considers the node-$k$'s rate constraint in~\eqref{con-rate-ul}. Hence, the AP's {receive} beamforming vector ${\bf w}_u$ can be aligned with the uplink channel ${\bf f}_{u,k}= {\bf h}^{{d}}_{k} + {\bf G}_{k} {\bm{\Theta}}_u {\bf h}^{{r}}_{k} $. Given ${\bf w}_u = {\bf f}_{u,k}/||{\bf f}_{u,k}||$, the optimal passive beamforming strategy $\bm{\Theta}_u$ needs to maximize the IRS-assisted channel gain $|{\bf f}_k^H{\bf w}_u|^2$ as follows:
\begin{equation}\label{prob-uplink-theta}
\max_{\theta_{u,n} \in (0, 2\pi]}\,\,||{\bf h}^{{d}}_{k} + {\bf G}_{k} {\bm{\Theta}}_u{\bf h}^{{r}}_{k}||^2,
\end{equation}
which can be easily solved by the SDR method similar to that in~\cite{irs-aoi-arxiv} and~\cite{SDR_our_work}. The transmission control parameters $(\tau_k, p_k)$ can be also optimized to minimize the energy consumption in~\eqref{obj-energy-ul}. Let $e_k \triangleq \tau_k p_k $ denote the node's energy consumption in RF communications. Given the optimal $\bm{\Theta}_u$ to~\eqref{prob-uplink-theta}, we can find the optimal $(\tau_k, p_k)$ by the following problem:
\begin{align}\label{prob-uplink-power}
\min_{e_k,\tau_k\in(0,1)} ~&~ e_k + p_c \tau_k, \quad \text{s.t.} \quad \tau_k \log\left(1+ \frac{e_k}{\tau_k}||{\bf f}^H_{u,k}||^2\right) \geq d_k. 
\end{align}
Problem~\eqref{prob-uplink-power} is convex in $(\tau_k,e_k)$ {and satisfies the Slater's condition, which allows us to find a closed-form solution by using the Lagrangian dual method~\cite{boyd2004}.
After this,} we can easily find the optimal transmit power as $p_k = e_k/\tau_k$. If the energy budget holds, i.e.,~$e_k + p_c \tau_k\leq E_k$, the node-$k$'s data transmission will be successful and thus we can update its AoI as $A_k(t+1) = 1$.

\subsection{Weighted Energy Transfer Maximization}

In downlink energy transfer with $\psi_{0}(t)= 1$, we aim to supply sufficient energy to all sensor nodes that can sustain their uplink information transmission to minimize the time-averaged AoI performance. However, it is difficult to explicitly quantify how downlink energy transfer affects the AoI performance. Instead, our intuitive design is to transfer more energy to those sensor nodes with the relatively worse AoI conditions. If the node-$k$ has a higher AoI value, we expect to transfer more energy to the node-$k$. This allows the node-$k$ to increase its {sampling} frequency and report its sensing information with a shorter transmission delay, and thus reducing its AoI value in the following sensing and reporting cycles. By this intuition, in the downlink phase we aim to maximize the AoI-weighted energy transfer to all sensor nodes, formulated as follows:
\begin{subequations}\label{prob-energy-dl}
\begin{align}
\max_{{\bf w}_d,\bm{\Theta}_d} ~&~ \sum_{k\in\mathcal{K}}v_{k}  | {\bf f}^H_{d,k}  {\bf w}_d |^2 \label{obj_max_energy}\\
\text{s.t.}
~&~ E^c_{k} \leq E_k  + E^h_k  ,\quad  \forall\,k\in\mathcal{K},\label{con-energy-dl}\\
~&~||{\bf w}_d || \leq 1 \text{ and } { \theta}_{d,n} \in (0, 2\pi), \quad \forall\, n\in\mathcal{N}, \label{theta-beam-dl}
\end{align}
\end{subequations}
where $E^c_{k} = \tau_k (p_k  + p_c)$ denotes the node-$k$'s energy consumption in the uplink transmission. For each node-$k$, we define the weight parameter as $v_{k} =A_k +\alpha_k E_k^{-1}$, {which} is increasing in the AoI value $A_k$ while inversely proportional to the energy capacity $E_k$. Thus, we prefer to transfer more energy to the sensor node with a higher AoI value and a lower energy supply, which is prioritized by a larger weight parameter $v_{k}$ in~\eqref{obj_max_energy}. Such a heuristic is expected to reduce the average AoI of all sensor nodes in a long term. The constant $\alpha_k$ characterizes the tradeoff between the sensor node's energy supply and AoI performance. A larger value of $\alpha_k$ indicates that the sensor node is more sensitive to the energy insufficiency. {Besides, we expect that any sensor node may need to upload its data in future time slots, but we do not know when it will be scheduled to transmit. For a conservative consideration, we impose the constraint~\eqref{con-energy-dl} to ensure that all sensor nodes will have sufficient energy to upload their data in the next time slot. If we remove~\eqref{con-energy-dl}, it becomes possible that some node-$k$ may not have sufficient energy to upload its data after beamforming optimization. If this node-$k$ happens to be scheduled by the DRL agent in the next time slot, its data transmission will be unsuccessful and thus its AoI will continue increasing at the AP.  Therefore, we include the constraint in~\eqref{con-energy-dl} as a one-step lookahead safety mechanism to ensure that every sensor has sufficient energy for data transmission when it is scheduled in the next time slot.}

\subsubsection{Alternating optimization (AO) for problem~\eqref{prob-energy-dl}}
Given the uplink $({\bf w}_u,\bm{\Theta}_u)$ and the control parameters $(\tau_k, p_k)_{k\in\mathcal{K}}$, the optimization of the downlink $({\bf w}_d,\bm{\Theta}_d)$ in~\eqref{prob-energy-dl} can be decomposed by the AO method into two sub-problems, {similar to that in~\cite{SDR_our_work}.} In the first sub-problem, we optimize $\bm{\Theta}_d$ in problem~\eqref{prob-energy-dl} with the fixed ${\bf w}_d$. Note that only the IRS-enhanced downlink channel ${\bf f}_{d,k}$ relates to $\bm{\Theta}_d$ as shown in~\eqref{channel_down}. We can simplify problem~\eqref{prob-energy-dl} by introducing a few auxiliary variables. Define ${\bf F}_{k}\triangleq {\bf G}_k \text{diag}({\bf h}^{{r}}_{k} )$ and then we can simplify~\eqref{channel_down} as ${\bf f}_{d,k} = {\bf h}^{{d}}_{k}(t) + {\bf F}_{k}{\bm{\theta}}_d$. We further define $\bar{\bm{\theta}}_d \triangleq [ \bm{\theta}_d, \zeta]^T$ where $\zeta \geq 0$ and $|\zeta|=1$. The quadratic term in~\eqref{obj_max_energy} can be rewritten as $|{\bf f}^H_{d,k} {\bf w}_d|^2 =\bar{\bm{\theta}}_d^H {\bf R}_k\bar{\bm{\theta}}_d + ({\bf h}^{{d}}_{k})^H{\bf h}^{{d}}_{k}$, where the matrix coefficient ${\bf R}_k$ is given by ${\bf R}_k = \begin{bmatrix}
{\bf F}^H_{k}{\bf w}_d{\bf w}_d^H{\bf F}_{k} &
{\bf F}^H_{k}{\bf w}_d{\bf w}_d^H{\bf h}^{{d}}_{k}
\\
\bigr({\bf h}^{{d}}_{k}\bigr)^H{\bf w}_d{\bf w}_d^H
{\bf F}_{k}&
0
\end{bmatrix}$. We can further apply SDR to $\bar{\bm{\theta}}^H_d {\bf R}_k \bar{\bm{\theta}}_d$ by introducing the matrix variable $\bm{\Phi}_d = \bar{\bm{\theta}}_d\bar{\bm{\theta}}^H_d$. Similar transformation can be applied to the energy budget constraint in~\eqref{con-energy-dl}. As such, the optimization of the downlink $\bm{\Theta}_d$ can be converted into the following SDP similar to that in~\cite{SDR_our_work} and~\cite{our_work_sdp}.
\begin{subequations}\label{prob-energy-dl-SDP}
\begin{align}
\max_{{\bm \Phi}_d \succeq 0} ~&~ \sum_{k\in\mathcal{K}}v_{k} \textbf{Tr}\bigl({\bf R}_k {\bm \Phi}_d\bigr) \label{obj_max_energy_SDP}\\
\text{s.t.}
~&~  \textbf{Tr}\bigl({\bf R}_k{\bm \Phi}_d\bigr) \geq (\eta p_s)^{-1}(E^c_{k}  - E_k) ,\quad \forall\, k\in\mathcal{K},\label{con-energy-dl-sdp}\\
~&~ {\bm \Phi}_d(n, n)=1, \quad \forall\, n\in\mathcal{N},
\end{align}
\end{subequations}
where $\textbf{Tr}(\cdot)$ denotes the matrix trace. Given the constant weight $v_{k}$, problem~\eqref{prob-energy-dl-SDP} becomes a conventional beamforming optimization for downlink MISO system~\cite{our_other_work}, which can be solved efficiently by the interior-point algorithm. In the second sub-problem, we optimize ${\bf w}$ in problem~\eqref{prob-energy-dl} with the fixed ${\bm{\Theta}}_d$ and $(\tau_k, p_k)$. This follows a similar SDR approach as that in~\eqref{prob-energy-dl-SDP} by introducing a matrix variable ${\bf W}_d = {\bf w}_d{\bf w}_d^H$. Once the optimal solution ${\bf W}_d$ or ${\bm{\Phi}}_d$ is obtained, we can extract the rank-one beamformer ${\bf w}_d$ or ${\bm\theta}_d$ by Gaussian randomization method. We continue the iterations between ${\bf w}_d$ and ${\bm{\Theta}}_d$ until the convergence to a stable point. 
\begin{algorithm}[t]
	\caption{AO Method for Downlink Energy Transfer}
	\label{alg-AO}
	\begin{algorithmic}[1]
	\renewcommand{\algorithmicrequire}{\textbf{Input:}}
	\REQUIRE The channel information $\{{\bf h}^{{d}}_{k}(t),{\bf G}_{k}(t),{\bf h}^{{r}}_{k}(t)\}$, AoI state $A_k$, energy state $E_k$, and energy demand $E_k^c$ of each sensor node $k\in\mathcal{K}$
	\renewcommand{\algorithmicrequire}{\textbf{Initialize:}}
	\REQUIRE a feasible beamforming strategy $({\bf w}_d,{\bm{\Theta}}_d)$, $\tau\leftarrow 0$
\STATE $v_{k} \leftarrow A_k +\alpha_k E_k^{-1}$
\STATE $E_d^{(\tau)}\leftarrow 0$, $E_d^{(\tau+1)}\leftarrow \sum_{k\in\mathcal{K}}v_{k}  | {\bf f}^H_{d,k}  {\bf w}_d |^2$
\WHILE{$|| E_d^{(\tau+1)} - E_d^{(\tau)} ||\geq \epsilon$}
    \STATE $\tau\leftarrow \tau +1 $
	\STATE Solve SDP~\eqref{prob-energy-dl-SDP} by the interior-point algorithm
	\STATE Extract the rank-one passive beamformer ${\bm{\Theta}}_d$
	\STATE Update ${\bf w}_d$ with the fixed $\bm{\Theta}_d$
	\STATE $E_d^{(\tau)}\leftarrow E_d^{(\tau+1)}$
    \STATE $E_d^{(\tau+1)}\leftarrow \sum_{k\in\mathcal{K}}v_{k}  | {\bf f}^H_{d,k}  {\bf w}_d |^2$
\ENDWHILE

	\end{algorithmic}
\end{algorithm}
\subsubsection{Simple approximation to problem~\eqref{prob-energy-dl}}
Given the scheduling decision $\psi_{k}(t)$, the inner-loop optimization estimates the beamforming strategies $({\bf w}_d, {\bm \Theta}_d)$ and the transmission parameters $(\tau_k,p_k)_{k\in\mathcal{K}}$. The inner-loop optimization should be very efficient to minimize the computational overhead and processing delay in each iteration. Note that the AO algorithm for $({\bf w}_d,\bm{\Theta}_d)$ can be inefficient as each iteration requires to solve the SDP problem~\eqref{prob-energy-dl-SDP} with a high computational complexity. The number of AO iterations till convergence is also unknown. Instead of the AO method, we further propose a simple approximation to problem~\eqref{prob-energy-dl} by optimizing $\bm{\Theta}_d$ with a fixed and feasible ${\bf w}_d$. {This solution} can avoid the iterations between ${\bf w}_d$ and $\bm{\Theta}_d$, and thus improve the {inner-loop computation efficiency}. In particular, we firstly consider an optimistic case in which the AP's downlink beamforming vector ${\bf w}_d$ can be aligned to all users' downlink channels ${\bf f}^H_{d,k}$, and thus we can relax problem~\eqref{prob-energy-dl} as follows:
\begin{equation}\label{prob-energy-dl-appx}
\max_{\bm{\Theta}_d}  \,\, \sum_{k\in\mathcal{K}}v_{k}  || {\bf f}_{d,k}||^2, \quad \text{s.t.} \quad \eqref{con-energy-dl}-\eqref{theta-beam-dl},
\end{equation}
which {only relies on ${\bm{\Theta}_d}$ and} can be solved by a similar SDR method for problem~\eqref{prob-uplink-theta}. However, problem~\eqref{prob-energy-dl-appx} overestimates the total energy transfer to all sensor nodes. In the second step, we can reorder the sensor nodes by the channel gain $||{\bf f}_{d,k}||^2$ and fix the downlink beamforming vector as ${\bf w}_d = {\bf f}^m_{d,k}/||{\bf f}^m_{d,k}||$, where ${\bf f}^m_{d,k} = \arg\min_{k\in\mathcal{K}} ||{\bf f}_{d,k}||^2$. This intuition ensures that we transfer more RF energy to the sensor node with the worst channel condition. In the third step, we optimize $\bm{\Theta}_d$ by solving~\eqref{prob-energy-dl-SDP} with the fixed ${\bf w}_d$, which provides a feasible lower bound to problem~\eqref{prob-energy-dl}. As such, we only need to solve two SDPs to approximate the solution $({\bf w}_d,\bm{\Theta}_d)$.
\begin{algorithm}[t]
	\caption{Simple Approximation for $({\bf w}_d,\bm{\Theta}_d)$}
	\label{alg-Simple}
	\begin{algorithmic}[1]
	\renewcommand{\algorithmicrequire}{\textbf{Input:}}
	\REQUIRE The channel information $\{{\bf h}^{{d}}_{k}(t),{\bf G}_{k}(t),{\bf h}^{{r}}_{k}(t)\}$, AoI state $A_k$, energy state $E_k$, and energy demand $E_k^c$ of each sensor node $k\in\mathcal{K}$
    \STATE $v_{k} \leftarrow A_k +\alpha_k E_k^{-1}$
    \STATE Solve problem~\eqref{prob-energy-dl-appx} in the optimistic case
    \STATE ${\bf f}^m_{d,k} \leftarrow \arg\min_{k\in\mathcal{K}} ||{\bf f}_{d,k}||^2$
    \STATE ${\bf w}_d \leftarrow {\bf f}^m_{d,k}/||{\bf f}^m_{d,k}||$
	\STATE Solve problem~\eqref{prob-energy-dl-SDP} with the fixed ${\bf w}_d$
	\STATE Extract the passive beamforming strategy ${\bm{\Theta}}_d$
	\end{algorithmic}
\end{algorithm}

\section{Outer-loop Learning for Scheduling}\label{sec-outerloop}

The outer-loop DRL approach aims to update the AP's scheduling policy by continuously interacting with the network environment. We can reformulate the scheduling optimization problem into the Markov decision process (MDP), which can be characterized by a tuple $(\mathcal{S},\mathcal{A},\mathcal{R})$. The state space $\mathcal{S}$ denotes the set of all system states. In each decision epoch, the AP's state ${\bf s}_t\in\mathcal{S}$ includes all sensor nodes' AoI values, denoted as a vector ${\bf A}(t) \triangleq [A_1(t),A_2(t),\ldots, A_K(t)]$, and the energy status denoted as ${\bf E}(t) = [E_1(t),E_2(t),\ldots,E_K(t)]$. Hence, we can define the system state in each time slot as ${\bf s}_t=({\bf A}(t), {\bf E}(t))$. {For each sensor node-$k$, we have $A_k(t)\geq 1$ as its AoI value can keep increasing from 1. The energy status $E_k(t)$ is upper bounded by the maximum battery capacity, i.e.,~$E_k(t)\in[0,E_{\max}]$.} The action space $\mathcal{A}$ denotes the set of all feasible scheduling decisions ${\bf a}_t \triangleq \{\psi_0(t), \psi_1(t),\ldots,\psi_K(t)\}\in\{0,1\}^{K+1}$ that satisfies the inequality in~\eqref{equ-sched}. Given the AP's scheduling decision, we can obtain $({\bf w}_d,\bm{\Theta}_d)$ by the inner-loop optimization and then update the AoI performance of different sensor nodes. The reward $\mathcal{R}$ assigns each state-action pair an immediate value $v_t({\bf s}_t, {\bf a}_t)$. It also {influences} the DRL agent's action adaptation to maximize the long-term reward, namely, the value function $V_{\pi}({\bf s}_0)\triangleq\sum_{t\in\mathcal{T}} \varepsilon^t v_t({\bf s}_t, {\bf a}_t)$, {where $\varepsilon\in(0,1)$ is} the discount factor for cumulating the reward and $\pi({\bf a}_t|{\bf s}_t)$ denotes the policy function mapping each state ${\bf s}_t$ to the action ${\bf a}_t$. Specifically, considering the AoI minimization in~\eqref{prob-aoi}, we can define the reward $v_t({\bf s}_t, {\bf a}_t)$ in the $t$-th time step as follows:
\begin{equation}\label{reward}
v_t({\bf s}_t, {\bf a}_t) = - \frac{1}{K|\mathcal{H}_t|}\sum_{\tau\in\mathcal{H}_t}\sum_{k\in\mathcal{K}} \lambda_k A_k(\tau),
\end{equation}
where $\mathcal{H}_t \triangleq \{t-t_o, \ldots, t-1, t\}\subset\mathcal{T}$ denotes a set of past time slots with the length $|\mathcal{H}_t|$. Hence, the reward $v_t({\bf s}_t, {\bf a}_t)$ {can be considered} as the averaged AoI of all sensor nodes in the most recent sliding window of the past time slots.

DRL approaches such as the value-based DQN and policy-based policy gradient (PG) algorithms have been demonstrated to be effective for solving MDP in large-scale wireless network by using DNNs to approximate either the value function $V_{\pi}({\bf s}_t)$ or the policy function $\pi({\bf a}_t|{\bf s}_t)$~\cite{network19}. In particular, the DQN method is an extension of the classic Q-learning method~\cite{drl-survey}, by using DNN to approximate the Q-value function $Q_{\bm \mu}({\bf s}_t,{\bf a}_t)$, where $\bm \mu$ denotes the DNN weight parameters for the Q-value network. Starting from the initial state ${\bf s}_0$, the PG algorithms directly optimize the value function $V_{\pi}({\bf s}_0)$ by using gradient-based approaches to update the policy $\pi_{\bm \omega}({\bf a}_t|{\bf s}_t)$, where $\bm\omega$ denotes the DNN weight parameters for the policy network. The proximal policy optimization (PPO) algorithm is recently proposed in~\cite{PPO} as a sample-efficient and easy-to-implement PG algorithm, striking a favorable balance between complexity, simplicity, and learning efficiency. In the sequel, we implement both the DQN and PPO algorithms and compare their performance for the outer-loop scheduling optimization.

\subsection{DQN Algorithm for Scheduling}

The DQN algorithm relies on two DNNs to stabilize the learning, denoted by the online Q-network and target Q-network. Given the DNN parameters ${\bm \mu}_t$ and ${\bm \mu}_t'$ for the two Q-networks, their outputs are given by $Q_{{\bm \mu}_t}({\bf s}_t, {\bf a}_t)$ and $Q_{{\bm \mu}_t'}({\bf s}_t, {\bf a}_t)$, respectively. At each learning episode, the DQN agent observes the system state ${\bf s}_t=({\bf A}(t), {\bf E}(t))$ and chooses the best scheduling action ${\bf a}_t$ with the maximum value of $Q_{{\bm \mu}_t}({\bf s}_t, {\bf a}_t)$. To enable random exploration, the DQN agent can also take a random action with a small probability. Once the action ${\bf a}_t$ is fixed and executed in the network, the DQN agent observes the instant reward $v_t({\bf s}_t, {\bf a}_t)$ and records the transition to the next state ${\bf s}_{t+1}$. Each transition sample $({\bf s}_t, {\bf a}_t, v_t, {\bf s}_{t+1})$ will be stored in the experience replay buffer. Meanwhile, the DQN agent estimates the target Q-value as $y_t = v_t({\bf s}_t, {\bf a}_t) + \varepsilon Q_{{\bm \mu}_t'}({\bf s}_{t+1}, {\bf a}_{t+1})$ by using the target Q-network with a different weight parameter ${\bm \mu}_t'$.

DQN's success lies in the design of the experience replay mechanism that improves the learning efficiency by reusing the historical transition samples to train the DNN in each iteration~\cite{network19}. The DNN training aims to adjust the parameter ${\bm \mu}_t$ to minimize a loss function $\ell({\bm \omega}_t)$, which is defined as the gap or more formally the temporal-difference (TD) error between the online Q-network $Q_{{\bm \mu}_t}({\bf s}_t, {\bf a}_t)$ and the target value $y_t$, specified as follows:
\begin{equation}\label{equ-loss-function}
\ell({\bm \mu}_t) = \mathbb{E}[|y_t - Q_{{\bm \mu}_t}({\bf s}_t, {\bf a}_t)|^2].
\end{equation}
The expectation in~\eqref{equ-loss-function} is taken over a random subset (i.e., mini-batch) of transition samples from the experience replay buffer. For each mini-batch sample, we can evaluate the target value $y_t$ and generate the Q-value $Q_{{\bm \mu}_t}({\bf s}_t, {\bf a}_t)$. The weight {parameters} ${\bm \mu}_t$ can be updated by {the backpropagation of the gradient information}. The DQN method stabilizes the learning by periodically copying the DNN parameter ${\bm \mu}_t$ of the online Q-network to the target Q-network.


\begin{figure}[t]
    \centering
    \includegraphics[width=0.9\textwidth]{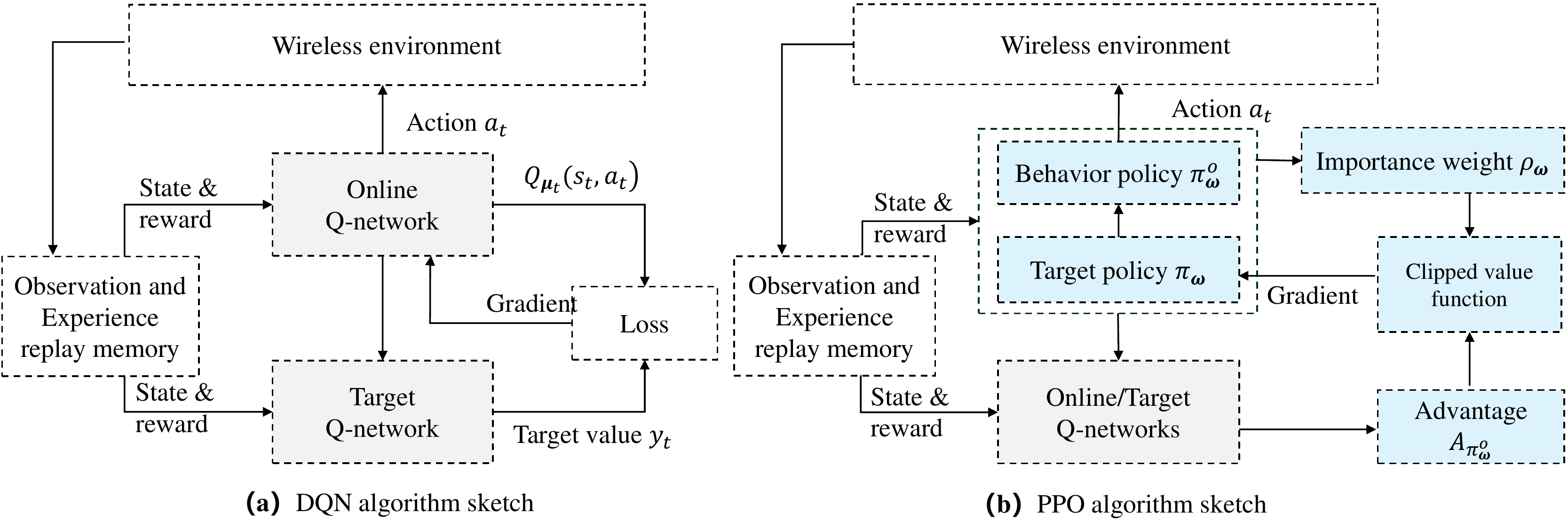}
    \caption{The DQN and PPO algorithms for the outer-loop scheduling optimization.}\label{fig-dqn-ppo}
\end{figure}

\subsection{Policy-based Actor-Critic Algorithms}

Different from the value-based DQN method, {the policy-based approach improves the value function by updating the parametric policy $\pi_{{\bm \omega}}$ in gradient-based methods}~\cite{network19}. Given the system state ${\bf s}_t$, the policy $\pi_{{\bm \omega}}({\bf a}_t|{\bf s}_t)$ specifies a probability distribution over different actions ${\bf a}_t\in\mathcal{A}$. The DNN training aims at updating the policy network to improve the expected value function, rewritten as $J({\bm \omega}) = \sum_{{\bf s} \in \mathcal{S}} d({\bf s}) V_{\pi}({\bf s}) = \sum_{{\bf s} \in \mathcal{S}} d({\bf s})  \sum_{{\bf a} \in \mathcal{A}} \pi_{\bm \omega}({\bf a} \vert {\bf s}) Q^{\pi}({\bf s},{\bf a})$,
where $d({\bf s})$ is the stationary state distribution corresponding to the policy $\pi_{\bm \omega}$ and $Q^{\pi}({\bf s},{\bf a})$ denotes the Q-value of the state-action pair $({\bf s},{\bf a})$ following the policy $\pi_{\bm \omega}$. Now the expected value function $J({\bm \omega})$ becomes a function of the policy parameter ${\bm\omega}$. The policy gradient theorem in~\cite{ddpg} simplifies the evaluation of the policy gradient $\nabla_{\bm \omega} J({\bm \omega})$ as follows:
\begin{equation}\label{equ-gpolicy}
\nabla_{\bm \omega} J({\bm \omega}) = \mathbb{E}_{\pi} \Big[ Q^{\pi}({\bf s},{\bf a}) \nabla_{\bm \omega} \ln\pi_{\bm \omega}({\bf a}|{\bf s})\Big],
\end{equation}
where the expectation is taken over all possible state-action pairs following the same policy $\pi_{\bm\omega}$.

For practical implementation, the policy gradient $\nabla_{\bm \omega} J({\bm \omega})$ can be evaluated by sampling the historical decision-making trajectories. At each learning epoch $t$, the DRL agent interacts with the environment by the state-action pair $({\bf s}_t, {\bf a}_t)$, collects an immediate reward $v_t$, and {observes} the transition to the next state ${\bf s}_{t+1}$. Let ${\bm \ell} \triangleq\{{\bf s}_1, {\bf a}_1, v_1, {\bf s}_2,{\bf a}_2, v_2, {\bf s}_3, \ldots, v_T\}$ denote the state-action trajectory as the DRL agent interacts with the environment. We can estimate the Q-value $Q^{\pi}({\bf s}_t,{\bf a}_t)$ in~\eqref{equ-gpolicy} by $g_t = \sum_{\tau=t}^T  \varepsilon^{\tau-t} v_t$. As such, we can approximate the policy gradient $\nabla_{\bm \omega} J({\bm \omega})$ in each time step by the random sample $g_t \nabla_{\bm \omega} \ln\pi_{\bm \omega}({\bf a}_t|{\bf s}_t)$ and update the policy network as ${\bm \omega}\leftarrow  {\bm \omega} + \alpha_{\bm\omega} g_t \nabla_{\bm \omega}\ln \pi_{\bm \omega}({\bf a_t} | {\bf s}_t)$, where $\alpha_{\bm\omega}$ denotes the step-size for gradient update. Besides the stochastic approximation, we can also employ another DNN to approximate the Q-value $Q^{\pi}({\bf s}_t,{\bf a}_t)$ in~\eqref{equ-gpolicy} and replace the Monte Carlo estimation $g_t$ by the DNN approximation $Q_{\bm \mu}({\bf s}_t,{\bf a_t})$ with the weight parameter~${\bm \mu}$, similar to that in the DQN algorithm. This motivates the actor-critic framework to update both the policy network and the Q-value network~\cite{ddpg}. The actor updates the policy network while the critic updates the Q-network by minimizing a loss function similar to~\eqref{equ-loss-function}. We can take the derivative of the loss function and update the weight parameter as ${\bm \mu}\leftarrow  {\bm \mu} + \alpha_{\bm\mu} \delta_t \nabla_{\bm \mu}Q_{\bm \mu}({\bf s}_t,{\bf a_t})$, where $\delta_t = y_t - Q_{{\bm \mu}_t}({\bf s}_t, {\bf a}_t)$ denotes the TD error. The Q-value estimation can be also replaced by the advantage function $A_{\pi}({\bf s}_t, {\bf a}_t) \triangleq Q^{{\pi}}({\bf s}_t, {\bf a}_t) - V_{\pi}({\bf s}_t)$ to reduce the variability in predictions and improve the learning efficiency.

The gradient estimation in~\eqref{equ-gpolicy} requires a complete trajectory by using the same target policy $\pi_{\bm \omega}$. It is actually the on-policy method that refrains us from using past experiences and limits the sample efficiency. This drawback can be avoided by a minor revision to the policy gradient:
\begin{equation}\label{equ-gpolicy-off}
\nabla_{\bm \omega} J({\bm \omega}) = \mathbb{E}_{\pi_{\bm\omega}^o} \Big[ \frac{\pi_{\bm \omega}({\bf s},{\bf a})}{\pi^{o}_{\bm \omega}({\bf s},{\bf a})}A_{\pi_{\bm\omega}}({\bf s},{\bf a}) \nabla_{\bm \omega} \ln\pi_{\bm \omega}({\bf a}|{\bf s})\Big],
\end{equation}
where the behavior policy $\pi_{\bm \omega}^o$ is used to collect the training samples. This becomes the off-policy gradient that allows us to estimate it by using the past experience collected from a different and even obsolete behavior policy $\pi_{\bm \omega}^o$. Hence, we can improve the sample efficiency by maintaining the experience replay buffer, similar to the DQN method. To further improve training stability, the off-policy trust region policy optimization (TRPO) method imposes an additional constraint on the gradient update~\cite{trpo}, i.e.,~the new policy $\pi_{\bm \omega}$ should not change too much from the old policy $\pi_{\bm \omega}^o$. Therefore, the policy optimization is to solve the following constrained problem:
\begin{align}\label{prob-trpo}
\max_{{\bm \omega}} ~&~ \mathbb{E}_{\pi_{\bm\omega}^o} \Big[ \frac{\pi_{\bm \omega}({\bf s},{\bf a})}{\pi^{o}_{\bm \omega}({\bf s},{\bf a})}A_{\pi_{\bm\omega}^o}({\bf s},{\bf a})\Big], \quad \text{s.t.} \quad \mathbb{E}_{\pi_{\bm\omega}^o}[ D_{KL}(\pi_{\bm\omega}, \pi_{\bm\omega}^o)]\leq \delta_{KL},
\end{align}
where $D_{KL}( P_1,P_2) \triangleq \int_{-\infty}^{\infty}P_1(x)\log\left(P_1(x)/P_2(x)\right)\,dx$ represents a distance measure in terms of the Kullback-Leibler (KL) divergence between two probability distributions~\cite{trpo}. The inequality constraint in~\eqref{prob-trpo} enforces that the KL divergence between two policies $\pi_{\bm\omega}$ and $\pi_{\bm\omega}^o$ are bounded by $\delta_{KL}$. The advantage function $A_{\pi_{\bm\omega}^o}$ in the objective of~\eqref{prob-trpo} is the approximation of the true advantage $A_{\pi_{\bm\omega}}$ corresponding to the target policy $\pi_{\bm\omega}$. However, the exact solution to the optimization problem~\eqref{prob-trpo} is not easy. Normally we require the first- and second-order approximations for both the objective and the constraint in~\eqref{prob-trpo}.

\subsection{PPO Algorithm for Scheduling}

The proximal policy optimization (PPO) algorithm proposed in~\cite{PPO} further improves the objective in~\eqref{prob-trpo} by limiting the probability ratio or the importance weight $\rho_{\bm \omega}({\bf s},{\bf a}) \triangleq \frac{\pi_{\bm \omega}({\bf s},{\bf a})}{\pi^{o}_{\bm \omega}({\bf s},{\bf a})}$ within a safer region $[1-\epsilon, 1+ \epsilon]$.
\begin{equation}\label{equ-ppo}
\tilde{J}(\pi_{\bm \omega}) = \mathbb{E}_{\pi_{\bm\omega}^o} \Big[ \min\Big\{\rho_{\bm \omega} A_{\pi_{\bm\omega}^o},\text{clip}(\rho_{\bm \omega}, 1-\epsilon, 1+\epsilon )A_{\pi_{\bm\omega}^o}\Big\}\Big],
\end{equation}
where the function $\text{clip}(\rho_{\bm \omega}, 1-\epsilon, 1+\epsilon)$ returns $\rho_{\bm \omega}$ if $\rho_{\bm \omega}\in[1-\epsilon, 1+ \epsilon]$ and returns $1-\epsilon$ (or $1+\epsilon$) if $\rho_{\bm \omega}<1-\epsilon$ (or $\rho_{\bm \omega}>1+\epsilon$). The parameter $\epsilon$ is used to control the clipping range. The approximate value function $\tilde{J}(\pi_{\bm \omega})$ ensures that the target policy ${\pi_{\bm\omega}}$ will not {deviate too} far from the behavior policy ${\pi_{\bm\omega}^o}$ for either positive or negative advantage $A_{\pi_{\bm\omega}^o}$.
With the clipped value function $\tilde{J}(\pi_{\bm \omega})$, we further introduce a Lagrangian dual variable $\beta_{KL}$ and reformulate the constrained problem~\eqref{prob-trpo} into the unconstrained maximization as follows:
\begin{equation}\label{prob-ppo}
\max_{{\bm \omega}} \,\, \tilde{J}(\pi_{\bm \omega}) - \beta_{KL} \mathbb{E}_{\pi_{\bm\omega}^o} \Big[   D_{KL}(\pi_{\bm\omega}, \pi_{\bm\omega}^o)\Big],
\end{equation}
The policy parameter ${\bm \omega}$ for the new value function in~\eqref{prob-ppo} can be easily updated by using the stochastic gradient ascent method. The parameter $\beta_{KL}$ can be also adapted according to the difference between the measured KL divergence $\mathbb{E}_{\pi_{\bm\omega}^o} \Big[   D_{KL}(\pi_{\bm\omega}, \pi_{\bm\omega}^o)\Big]$ and its target $\delta_{KL}$.

\begin{algorithm}[t]
	\renewcommand{\algorithmicrequire}{\textbf{Initialize:}}
	\caption{Energy-and-AoI-Aware Scheduling and Transmission Control Algorithm}
	\label{alg-ppo}
	\begin{algorithmic}[1]
	\REQUIRE Target policy $\pi_{\bm \omega}$ and behavior policy $\pi_{\bm \omega}^o$,
	\REQUIRE $t\leftarrow 0$, $E_k(0)\leftarrow B$, $A_k(t)\leftarrow 0$
	\FOR{$\text{Episode}=\{1, 2, \dots, \text{MAX}=3000\}$}
		\WHILE{$t \neq T$}
		\STATE Observe the system state $({\bf A}(t), {\bf E}(t))$
		\STATE Choose the outer-loop action $\bm{\Psi}(t)$ for scheduling
        \STATE {\bf case} $\psi_0(t)=0$: optimize uplink data transmission in~\eqref{prob-uplink-energy}
        \STATE {\bf case} $\psi_0(t)=1$: optimize downlink energy transfer in~\eqref{prob-energy-dl}
        \STATE Execute joint action ${\bf a}(t)\triangleq (\bm{\Psi}(t), {\bf w}(t),\bm{\Phi}(t))$, evaluate $v_t({\bf s}_{t},{\bf a}_{t})$
		\STATE Record the next state ${\bf s}_{t+1}$ and buffer the transition $({\bf s}_{t}, {\bf a}_{t}, v_{t}, {\bf s}_{t+1})$
		\STATE $t \gets t+1$
\ENDWHILE
	\STATE Take mini-batch form the experience replay buffer
	\STATE Estimate advantage $A_{\pi_{\bm\omega}^o}$ and update ${\bm \omega}$ to maximize~\eqref{prob-ppo}
	\STATE Update behavior policy $\pi_{\bm \omega}^o \gets (1-\mu)\pi_{\bm \omega}^o + \mu \pi_{\bm \omega}$
\ENDFOR
	\end{algorithmic}
\end{algorithm}

{As shown in Fig.~\ref{fig-dqn-ppo}, we highlight the comparison between the DQN and the PPO algorithms for outer-loop scheduling optimization. Different from the DQN algorithm, the PPO algorithm employs the actor-critic structure that relies on two sets of DNNs to approximate the policy networks. The difference between the target policy network and the behavior policy network is used to generate the importance sampling weight $\rho_{\bm \omega}({\bf s},{\bf a})$. The behavior policy network is used to interact with the environment and store the transition samples in the experience replay buffer. The target policy network is used to update the DNN weight parameter ${\bm\omega}$ by sampling a mini-batch randomly from the experience replay buffer. Then we can update the target policy network by maximizing the objective in~\eqref{prob-ppo}.} The complete solution procedures are listed in Algorithm~\ref{alg-ppo}. Considering the preferable learning efficiency and robustness of the PPO algorithm, we integrate it in Algorithm~\ref{alg-ppo} to adapt the outer-loop scheduling policy in the hierarchical learning framework. At the initialization stage, we randomly initialize the DNN weight parameters ${\bm\omega}$ for the policy network. In each learning episode, the AP collects observations $({\bf A}(t),{\bf E}(t))$ of the system and decides the outer-loop scheduling decision $\bm{\Psi}(t)$, as shown in line $4$ of Algorithm~\ref{alg-ppo}. Given the scheduling decision $\bf{\Psi(t)}$, the AP needs to optimize the joint beamforming strategies $({\bf w}(t),\bm{\Theta}(t))$ for either uplink information transmission or downlink energy transfer, corresponding to lines 5-6 of Algorithm~\ref{alg-ppo}. Note that the solution to downlink energy transfer can be determined by either the iterative Algorithm~\ref{alg-AO} or the simplified Algorithm~\ref{alg-Simple}. When we determine both the outer-loop and inner-loop decision variables, we can execute the joint action $(\bm{\Psi}(t), {\bf w}(t),\bm{\Theta}(t))$ in the wireless system and evaluate the reward function as shown in lines 7-8 of Algorithm~\ref{alg-ppo}. The DNN training is based on the random mini-batch sampled from the experience replay buffer, as shown in lines 11-13 of Algorithm~\ref{alg-ppo}. For performance comparison, the DQN algorithm is also implemented for outer-loop scheduling. Our numerical evaluation in Section~\ref{sec:num} reveals that the PPO-based algorithm can improve the convergence performance and achieve a lower AoI value compared to the DQN-based algorithm.

\begin{table}[]
    \centering
    \caption{Parameter settings}
    \setlength{\tabcolsep}{2.8mm}{
    \begin{tabular}{|c|c||c|c|}
    \hline
    \makebox[0.25\textwidth][c]{Parameters} & \makebox[0.12\textwidth][c]{Values} & \makebox[0.25\textwidth][c]{Parameters} & \makebox[0.12\textwidth][c]{Values} \\
    \hline
    Number of DNN hidden layers & $3$ & Optimizer & Adam\\
    \hline
    Actor's learning rate & $0.0005$ & Number of AP's antennas  & 4\\
    \hline
    Critic's learning rate & $0.001$ & AP's transmit power $p_s$ & $30$dBm\\
    \hline
    Reward discount & $0.99$ & Energy conversion efficiency $\eta$ & $0.9$\\
    \hline
    Number of neurons & $64$ & Noise power $\sigma^2$ & $-75$dBm\\
    \hline
    Activation function & Tanh and Softmax & Sensor nodes' data size $D$ & $5$Kbits\\
    \hline
    \end{tabular}}
    \label{system parameters}
    \vspace{-0.7cm}
\end{table}

\section{Simulation Results}\label{sec:num}
In this section, we present simulation results to verify the performance gain of the proposed algorithms for the wireless-powered and IRS-assisted wireless sensor network.
The $(x, y, z)$-coordinates of the AP and the IRS in meters are given by $(200,-200,0)$ and $(0,0,0)$, respectively. The sensor nodes are randomly distributed in a rectangular area $[5,35] \times [-35, 35]$ in the $(x, y)$-plane with $z=-20$. We consider that the direct channel from the AP to each sensor node-$k$ follows the Rayleigh fading distribution, i.e., ${\bf h}^{{d}}_{k}(t) = \beta_{0, k} \tilde{\bf h}^{{d}}_{k}(t)$, where $\tilde{\bf h}^{{d}}_{k}(t) \thicksim \mathcal{CN}(0, {\bm I})$ denotes the complex Gaussian random variable and $\beta_{0, k}$ denotes the path-loss of the direct channel. Given the distance $d^{\rm{AS}}_k$ from the AP to the sensor node-$k$, we have $\beta_{0, k} = 32.6 + 36.7\log(d^{\rm{AS}}_k)$. A similar channel model is employed in~\cite{path_loss_set_2}. The IRS-sensor channel ${\bf h}^{{r}}_{k}(t)$ and the AP-IRS channel ${\bf G}_{k}(t)$ are modelled similarly. More detailed parameters are summarized in Table~\ref{system parameters}.


\begin{figure}[t]
    \centering
    \subfloat[Reward dynamics in outer-loop learning algorithms]{\includegraphics[width=\doublesize\textwidth]{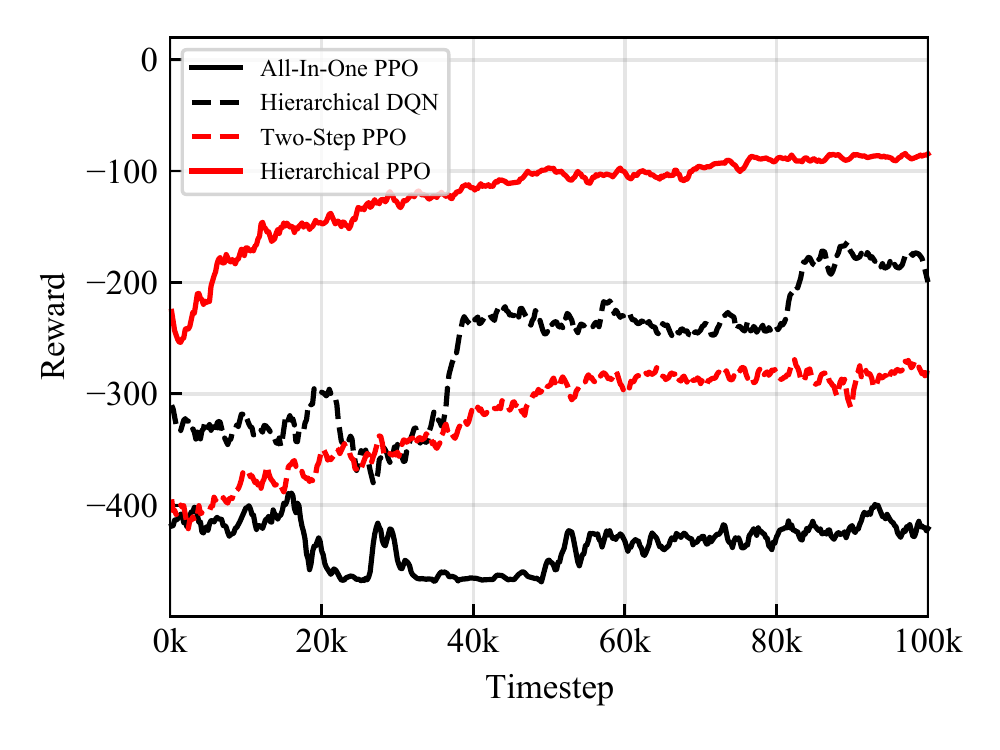}}
    \subfloat[Convergence with linear and non-linear EH models]{\includegraphics[width=\doublesize\textwidth]{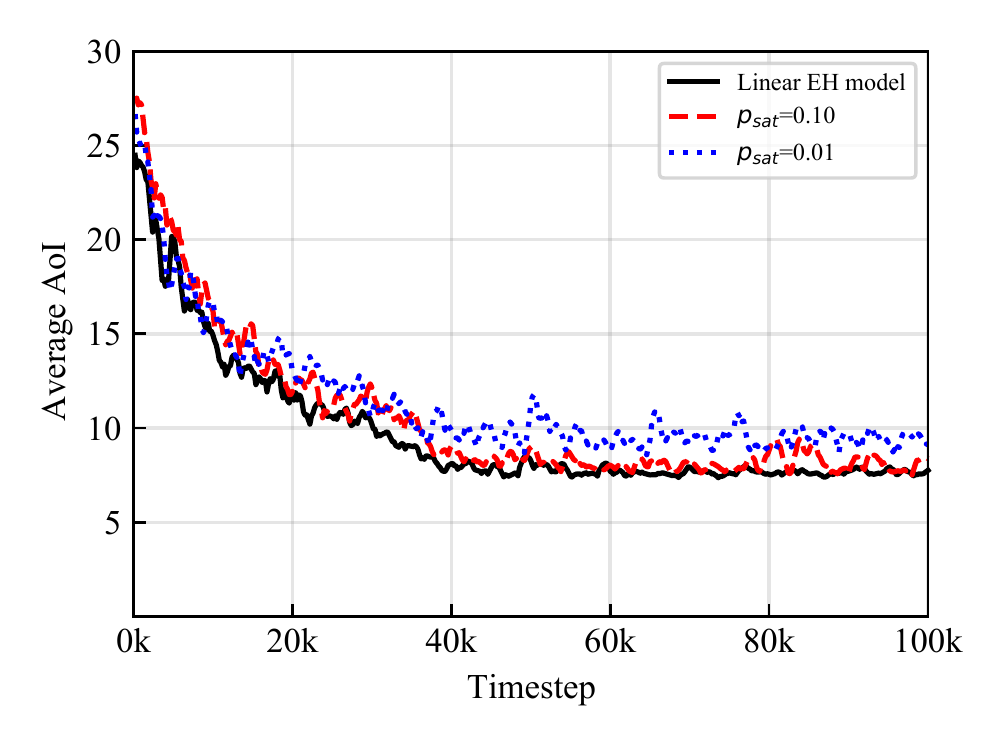}}\\
    \subfloat[Run time overhead for the inner-loop optimization]{\includegraphics[width=\doublesize\textwidth]{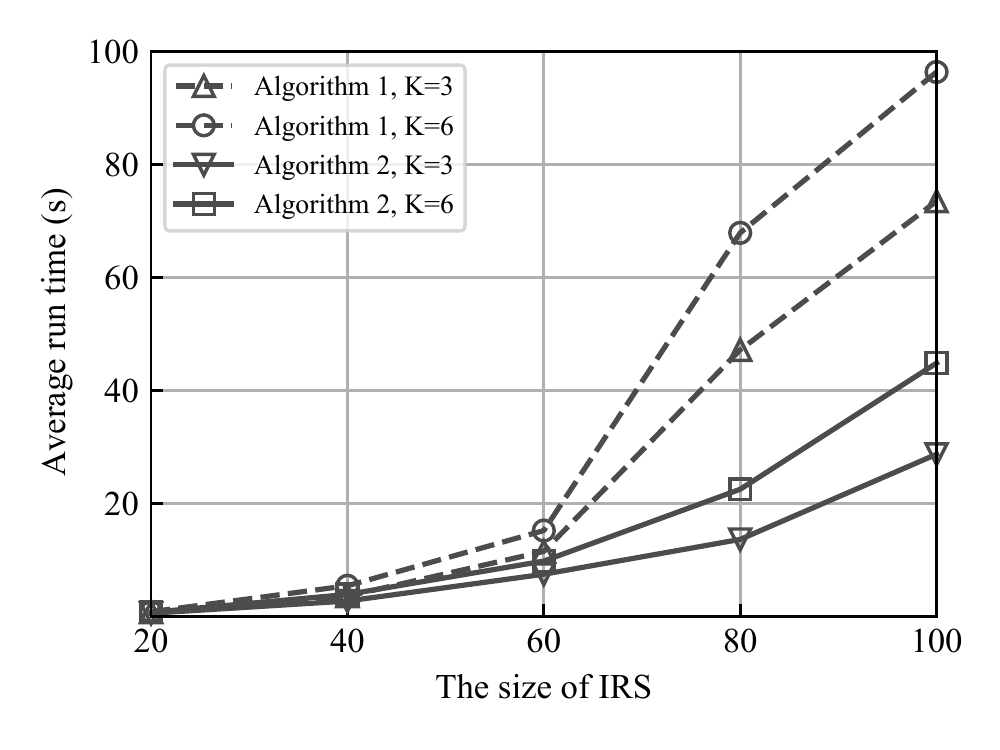}}
    \subfloat[Reward dynamics for the inner-loop optimization]{\includegraphics[width=\doublesize\textwidth]{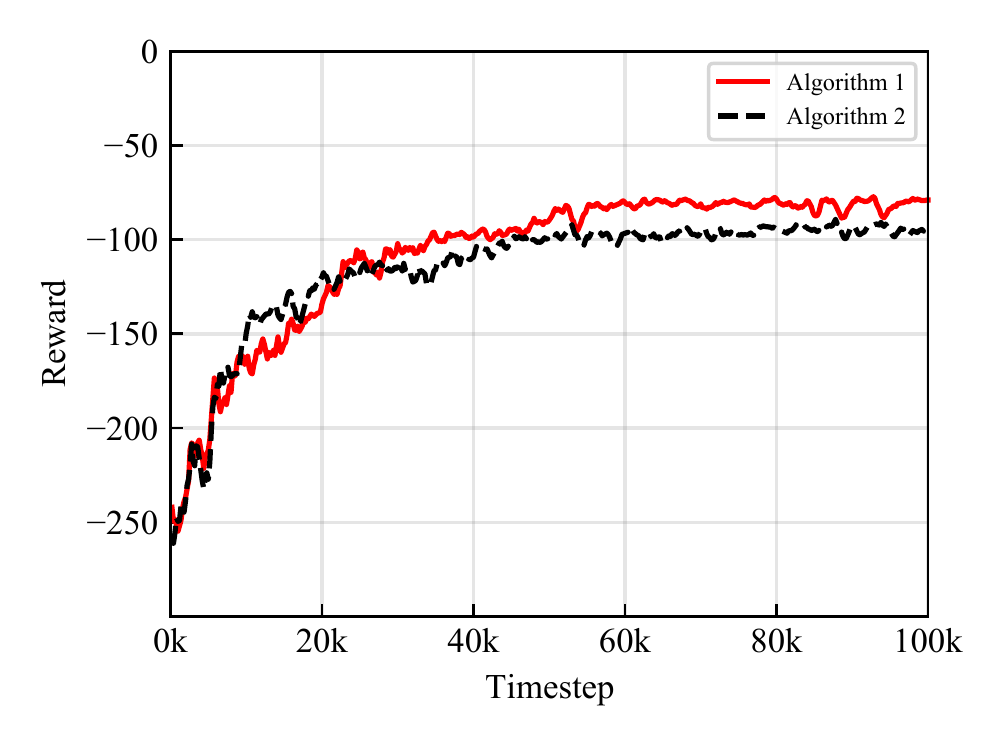}}
    \caption{The outer-loop learning and inner-loop optimization of the hierarchical PPO algorithm.}\label{fig-reward}
    \vspace{-1cm}
\end{figure}

\subsection{Improved Learning Efficiency and Convergence}

In Fig.~\ref{fig-reward}(a), we compare the reward performance among the hierarchical learning Algorithm~\ref{alg-ppo}, the hierarchical DQN and the conventional PPO, in which all decision variables $(\bm{\Psi}(t), {\bf w}(t),\bm{\Theta}(t))$ are adapted simultaneously. Hence, we denote the conventional PPO as the All-in-One PPO algorithm in Fig.~\ref{fig-reward}. The solid line in red represents the dynamics of the AoI performance in the proposed hierarchical PPO algorithm, which spits the decision variables into two parts and optimizes the beamforming strategy $({\bf w}(t),\bm{\Theta}(t))$ by the inner-loop Algorithm~\ref{alg-AO}. The dotted line in red denotes hierarchical DQN algorithm, and the dash-dotted line in black denotes the conventional All-in-One PPO algorithm. It is clear that the All-in-One PPO may not converge effectively due to a huge action space in the mixed discrete and continuous domain. Compared with the PPO algorithm, the hierarchical DQN algorithm becomes {unstable} as shown in Fig.~\ref{fig-reward}. The hierarchical PPO for joint scheduling and transmission control can reduce the action space in the outer-loop learning framework and thus achieve a significantly higher reward performance and faster convergence guided by the inner-loop optimization modules. {We also implement the two-step learning algorithm (denoted as Two-Step PPO in Fig.~\ref{fig-reward}) in which both inner-loop and outer-loop control variables are adapted by the PPO learning methods. The Two-Step PPO algorithm has a better convergence than that of the conventional All-in-One PPO algorithm. However, its reward performance is much inferior to that of the hierarchical learning framework, which is guided by the inner-loop optimization-driven target during the outer-loop learning.}

Figure~\ref{fig-reward}(b) reveals how different EH models effect the AoI performance. The linear EH model results in a slightly smaller AoI value than that of the non-linear EH model. {The reason is that} the linear EH model over-estimates the sensor nodes' EH capabilities. For the non-linear EH model, the harvested power will not further increase as the received signal power becomes higher than the saturation power $p_{\text{sat}}$. Fig.~\ref{fig-reward}(b) shows that $p_{\text{sat}}$ also affects the AoI performance. Generally we can expect a better AoI performance with a higher saturation power $p_{\text{sat}}$. {In our simulation, we also evaluate the overall reward and average AoI performances with different discount factor $\varepsilon\in\{0.99, 0.95, 0.90, 0.80\}$, which is used to accumulate the rewards in different decision epochs. The simulation results reveal that the learning with a larger discount factor becomes stable. We further show the} run time and performance comparison between the iterative Algorithm~\ref{alg-AO} and and the simplified Algorithm~\ref{alg-Simple} for the inner-loop beamforming optimization. It is clear that the run time of each inner-loop optimization algorithm increases with the size of the IRS and the number of sensor nodes. Besides, we observe that Algorithm~\ref{alg-Simple} significantly saves the run time by reducing the number of iterations, especially with a large-size IRS, as shown in Fig.~\ref{fig-reward}(c). However, the reward performances of two algorithms are very close to each other, as shown in Fig~\ref{fig-reward}(d). This implies that we can deploy Algorithm~\ref{alg-Simple} preferably in practice.

%

\subsection{Performance Gain over Existing Scheduling Policies}

\begin{figure}[t]
    \centering
    \subfloat[Minimum AoI values compared with the baselines]{\includegraphics[width=\doublesize\textwidth]{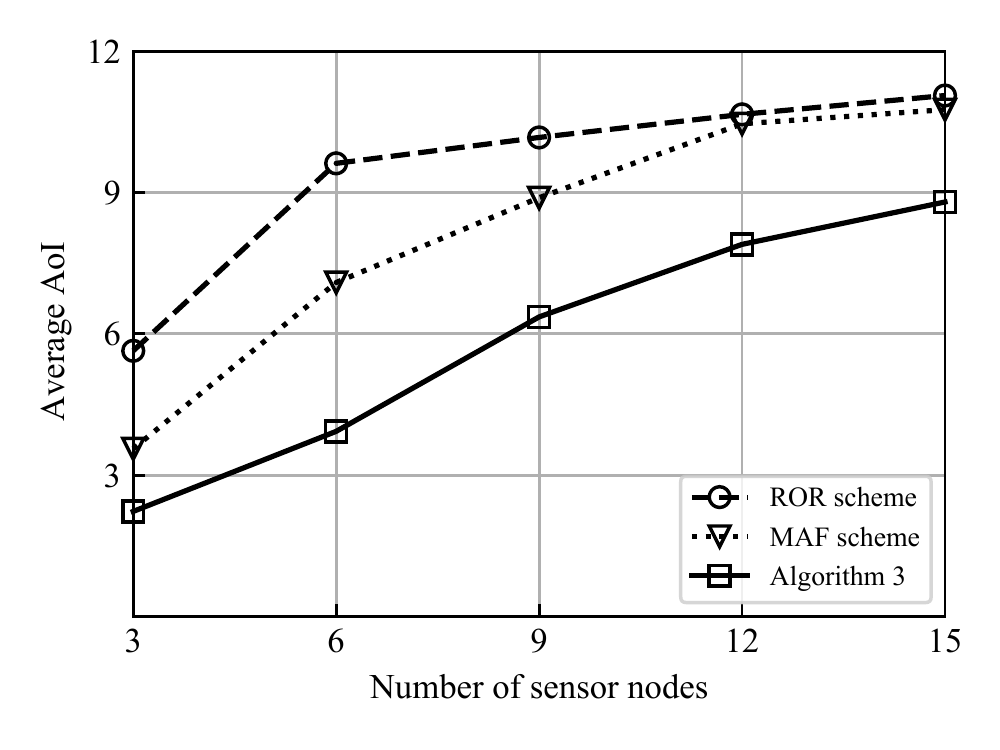}}
    \subfloat[Enhanced fairness among different nodes]{\includegraphics[width=\doublesize\textwidth]{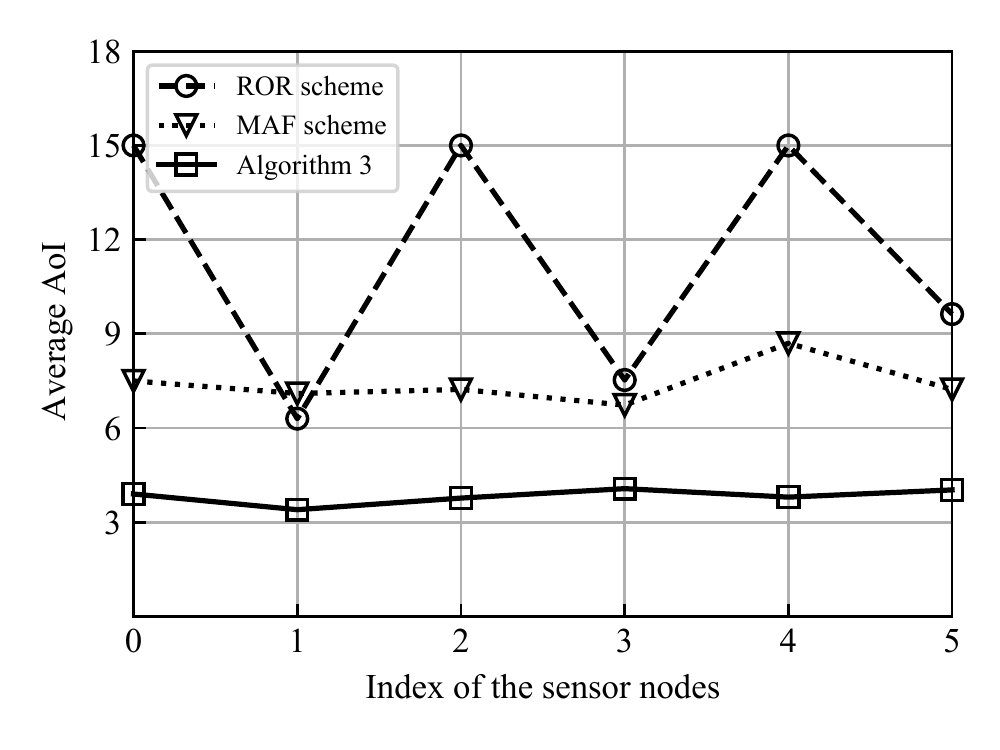}}
    \caption{Performance gain over baseline scheduling policies.}\label{fig-gain}
    \vspace{-1cm}
\end{figure}

In this part, we develop two baselines policies to verify the performance gain {of Algorithm~\ref{alg-ppo}}. The first baseline is the round-robin (ROR) scheduling policy which periodically selects one sensor node to upload its status-update information. In each scheduling period, we jointly optimize the active and passive beamforming strategy to enhance the information transmission. The second baseline is the Max-Age-First (MAF) scheduling policy, i.e., the AP selects the sensor node with the highest AoI value to upload its sensing information~\cite{maf18}. Both baselines rely on the same EH policy, i.e., the AP starts downlink energy transfer only when the scheduled sensor node has insufficient energy capacity, e.g., below some threshold value. In Fig.~\ref{fig-gain}(a), we show the AoI {performance as we increase the number of} sensor nodes. For different algorithms, we set the same coordinates for the AP and the IRS. Generally, different scheduling policies have a small AoI value {with a few sensor nodes}. The MAF policy performs better than the ROR policy, as it gives {higher priorities} to the sensor nodes with unsatisfactory AoI performance. As the number of nodes increases, Algorithm~\ref{alg-ppo} always outperforms the baselines by adapting the scheduling strategy according to sensor nodes' stochastic data arrivals, as shown in Fig.~\ref{fig-gain}(a).

In Fig.~\ref{fig-gain}(b), we compare the fairness of scheduling by showing the average AoI of different sensor nodes. For a fair comparison, we set the same weight parameters $\lambda_k$ for all sensor nodes in~\eqref{equ-waoi}. It can be seen from Fig.~\ref{fig-gain}(b) that Algorithm~\ref{alg-ppo} achieves a smaller AoI value for each sensor node. Moreover, different sensor nodes can achieve very similar AoI values, which implies the enhanced fairness in the scheduling policy by using Algorithm~\ref{alg-ppo}. {The ROR scheme has a large deviation compared to that of the other two baselines. The reason is that the RoR scheme cannot adapt to the dynamic data arrival process. An interesting observation is that the MAF scheme also has a smaller AoI deviation among different sensor nodes. This is because that the MAF scheme always chooses the sensor node with the highest AoI value to upload its sensing information. This can effectively prevent the AoI of some sensor nodes from being too high. }

\begin{figure}[t]
    \centering
    \subfloat[Larger-size IRS reduces transmission failure rate]{\includegraphics[width=\doublesize\textwidth]{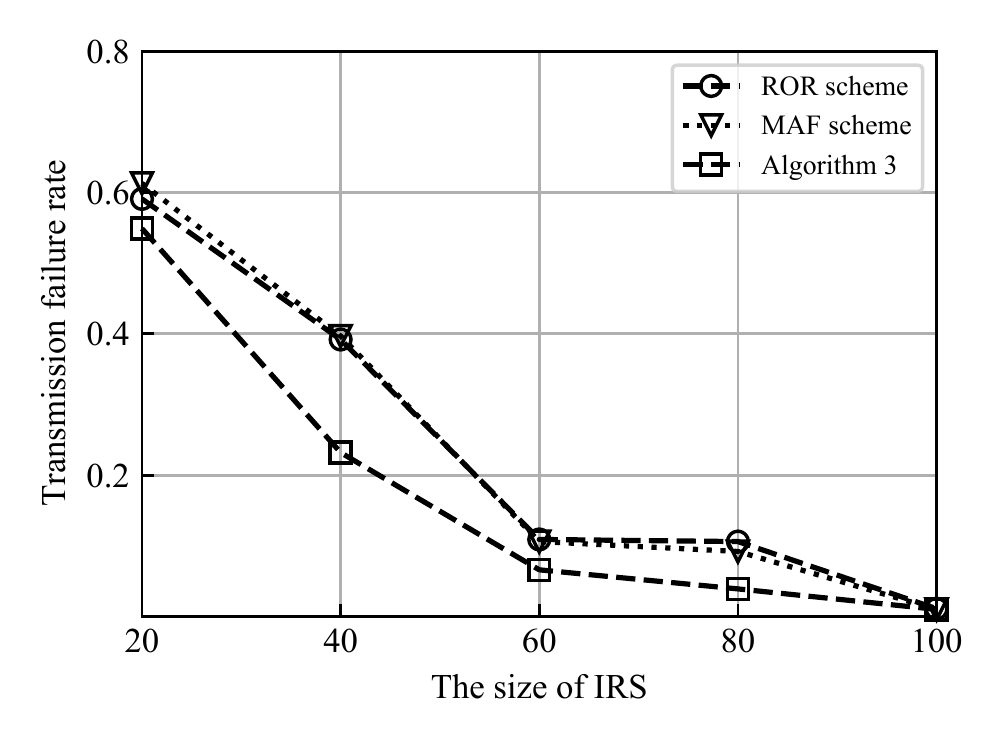}}
    \subfloat[Larger-size IRS improves the AoI performance.]{\includegraphics[width=\doublesize\textwidth]{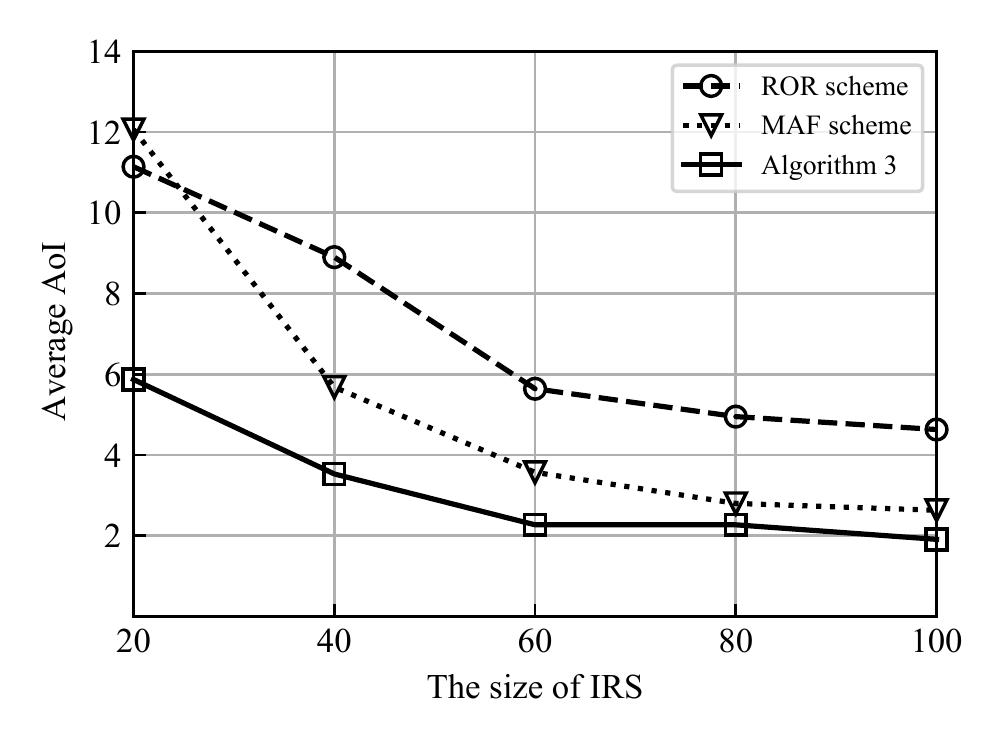}}
    \caption{Performance gain by using a larger-size IRS.}\label{fig-size}
    \vspace{-1cm}
\end{figure}

\subsection{Performance Gain by Using a Larger Size IRS}

{The use of IRS not only improves the downlink wireless energy transfer, but also enhances} the uplink channels for sensing information transmissions. Both aspects implicitly improve the {system's AoI performance. In this part}, we intend to verify the performance gain achievable by using the IRS. In Fig.~\ref{fig-size}, we show the dynamics of the sensor nodes' transmission failure rate and the average AoI by increasing the size of IRS from $20$ to $100$. Specifically, we count the number of transmission failures within $30$K time slots and {visualize} the transmission failure rate in Fig.~\ref{fig-size}(a). It is quite intuitive that a larger-size IRS can reduce the {sensor nodes' transmission failure rate, due to the IRS's reconfigurability} to improve the channel quality. The increase in {the IRS's size} makes it more flexible to reshape the wireless channels and improve the uplink transmission success probability. This can help minimize the AoI values in a long run. However, the AoI values will not keep decreasing as {the IRS's size} increases. As shown in Fig.~\ref{fig-size}(b), the performance gap between our method and the baselines becomes smaller as the size increases. That is because the channel conditions {become} much better with a large-size IRS and thus the bottleneck of AoI performance becomes the scheduling delay, instead of the transmission delay.


\begin{figure}[t]
    \centering
    \subfloat[Trade-off between AoI and energy consumption]{\includegraphics[width=\doublesize\textwidth]{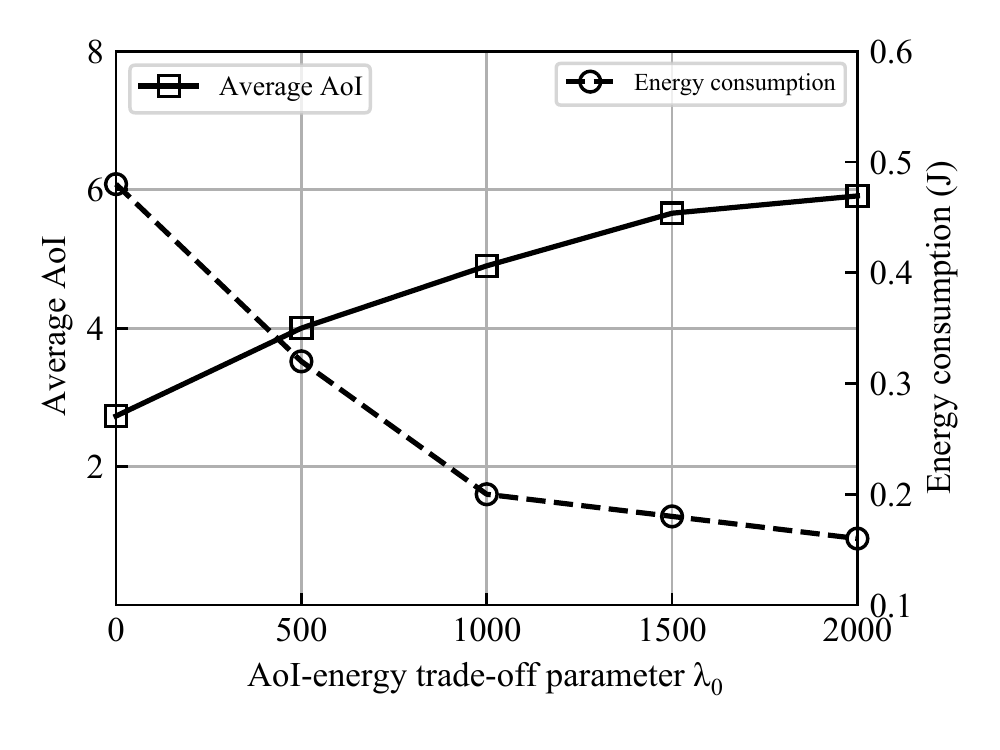}}
    \subfloat[Transmission scheduling with different priorities]{\includegraphics[width=\doublesize\textwidth]{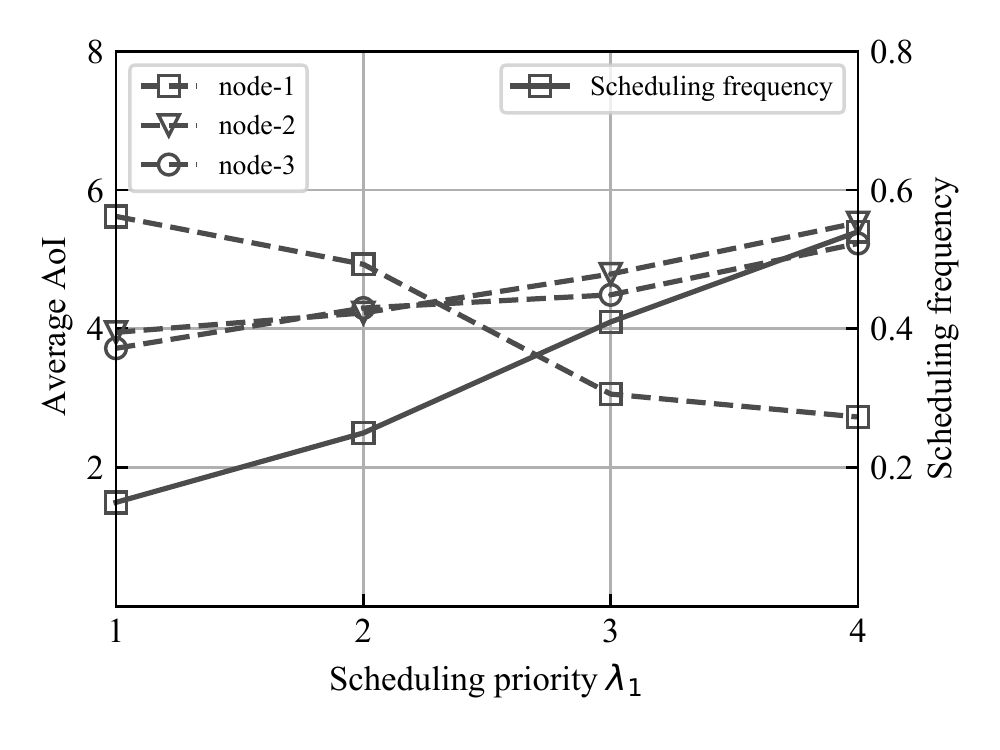}}
    \caption{Energy-aware AoI minimization with different user priorities. {We set $N=80$ and $K=3$ in the experiment}. }\label{fig-lambda}
    \vspace{-1cm}
\end{figure}

\subsection{Trade-off Between AoI and Energy Consumption}

In this part, we study the trade-off between AoI and energy consumption of the sensor nodes. Considering the AoI-energy tradeoff, we {can revise} the DRL agent's reward as follows:
\[
v_t({\bf s}_t, {\bf a}_t) = - \frac{1}{K|\mathcal{H}_t|}\sum_{\tau\in\mathcal{H}_t}\sum_{k\in\mathcal{K}} \Bigl( \lambda_k A_k(\tau) + \lambda_0 E^c_k(\tau) \Bigr),
\]
where the AoI-energy trade-off parameter $\lambda_0$ is used to balance the sensor node's AoI and energy consumption. With a smaller $\lambda_0$, the sensor node becomes more aggressive to upload its information. This may cause energy outage due to insufficient energy supply to the sensor node. With a larger $\lambda_0$, the sensor node will focus more on its energy status and {become} more tolerant to the information delay. As shown in Fig.~\ref{fig-lambda}(a), given different $\lambda_0$, the AoI value and energy consumption have different trends of evolution. When $\lambda_0 = 0$, the average AoI can be reduced by $53.8\%$ comparing to that with a higher $\lambda_0=2000$. We also show the performance gain with different priorities $\lambda_k$ for the sensor nodes. Considering three sensor nodes, we gradually increase $\lambda_1$ from 1 to 4 for the node-$1$ while setting fixed values for $\lambda_2=\lambda_3 = 2$. We evaluate the scheduling frequency in $30$K time slots and plot in Fig.~\ref{fig-lambda}(b) the change of node-$k$'s scheduling frequency, which is shown to increase linearly with respect to its priority $\lambda_1$. Fig.~\ref{fig-lambda}(b) also shows the change of three sensor nodes' AoI performance as we increase $\lambda_1$ for the node-$1$. It is clear to see that the node-$1$ will experience a much higher AoI value when it has a smaller priority, e.g.,~$\lambda_1 = 1$, than those of the other two nodes. When we gradually increase its priority, our algorithm will be more sensitive to the node-$1$'s AoI performance and thus try to reduce its AoI by scheduling it more often, as revealed in Fig.~\ref{fig-lambda}(b). As such, the node-$1$ will take up more transmission opportunities by sacrificing the AoI performance of the other two nodes. This verifies that our algorithm can be adaptive to the change of sensor nodes' priorities.



\section{Conclusions}\label{sec:cons}
In this paper, we have focused on a wireless-powered and IRS-assisted network and aimed to minimize the overall AoI for information updates. We have formulated the AoI minimization problem as a mixed-integer program and devised a hierarchical learning framework, which includes the outer-loop model-free learning and the inner-loop optimization methods. {Simulation results demonstrate that our algorithm can significantly reduce the average AoI and achieve controllable fairness among sensor nodes. More specifically, the hierarchical PPO algorithm achieves a significantly higher reward performance and faster convergence compared to the hierarchical DQN algorithm. It also outperforms typical baseline strategies in terms of the AoI performance and fairness. The performance gain can be more significant with a small size IRS.}

\vspace{-0.5cm}
\bibliographystyle{IEEEtran}

\bibliography{reference}{}

\end{document}